\newcommand{\Teff}{T$_{\mathrm{eff}}$ }
\newcommand{\Teffnospace}{T$_{\mathrm{eff}}$}
\newcommand{\kms}{km~s$^{-1}$ }
\newcommand{\kmsnospace}{km~s$^{-1}$}
\newcommand{\afe}{[$\alpha$/Fe] }

\newcommand{\feh}{[Fe/H]}
\newcommand{\afenospace}{[$\alpha$/Fe]}
\newcommand{\apoor}{$\alpha$-poor }
\newcommand{\arich}{$\alpha$-rich }
\newcommand{\apoornospace}{$\alpha$-poor}
\newcommand{\arichnospace}{$\alpha$-rich}
\documentclass[useAMS,usenatbib]{mn2e}
\usepackage{graphicx}
\usepackage{amssymb}
\usepackage{times}
\usepackage{subfigure}
\usepackage[T1]{fontenc}
\usepackage{aecompl}
\usepackage[breaklinks,colorlinks,citecolor=blue]{hyperref}

\begin{document}

\title[APOGEE Galactic disk/halo Interface]{Using chemical tagging to redefine the interface of the Galactic disk and halo}
 \author[K. Hawkins et. al.]{K. Hawkins$^{1}$\thanks{E-mail: khawkins@ast.cam.ac.uk}, P. Jofr\'e $^{1}$, T. Masseron$^{1}$, and G. Gilmore$^{1}$  \\
$^{1}$Institute of Astronomy, Madingley Road, Cambridge. CB3 0HA}

\date{Accepted 2015 July 13.  Received 2015 July 13; in original form 2015 May 16}


\maketitle

\label{firstpage}

\begin{abstract}
We present a chemical abundance distribution study in 14 $\alpha$, odd-Z, even-Z, light, and Fe-peak elements of approximately 3200 intermediate metallicity giant stars from the APOGEE survey. The main aim of our analysis is to explore the Galactic disk-halo transition region between --1.20~$<$\feh\ $<$~--0.55 as a means to study chemical difference (and similarities) between these components. In this paper, we show that there is an \apoor and \arich sequence within both the metal-poor and intermediate metallicity regions. Using the Galactic rest-frame radial velocity and spatial positions, we further separate our sample into the canonical Galactic components. We then studied the abundances ratios, of Mg, Ti, Si, Ca, O, S, Al, C+N, Na, Ni, Mn, V, and K for each of the components and found the following: (1) the \apoor halo subgroup is chemically distinct in the $\alpha$-elements (particularly O, Mg, and S), Al, C+N, and Ni from the \arich halo, consistent with the literature confirming the existence of an \apoor accreted halo population; (2) the canonical thick disk and halo are not chemically distinct in all elements indicating a smooth transition between the thick disk and halo; (3) a subsample of the \apoor stars at metallicities as low as [Fe/H] $\sim$ --0.85 dex are chemically and dynamically consistent with the thin disk indicating that the thin disk may extend to lower metallicities than previously thought, and (4) that the location of the most metal-poor thin disk stars are consistent with a negative radial metallicity gradient. Finally, we used our analysis to suggest a new set of chemical abundance planes (\afenospace, [C+N/Fe], [Al/Fe], and [Mg/Mn]) that may be able to chemically label the Galactic components in a clean and efficient way independent of kinematics. 
\end{abstract}

\section{Introduction}
\label{sec:Introduction}
The stars in the Milky Way change their properties over time in almost every aspect. On one hand, their atmospheric parameters change as they age, producing significant changes in their colours and luminosities, with complete generations of stars enriching the interstellar medium with chemical elements as they die and new generations forming from this enriched material. On the other hand, their positions and kinematics change as they move with orbits governed mostly by gravitation. The disentanglement of their spatial and dynamical properties, complemented with their ages and chemical compositions (when possible), are the tools that have been used for decades to classify the stars into different stellar populations building up a picture of the structure of the Milky Way we have today \citep[e.g.][and references therein]{Gilmore1989, Freeman2002, Helmi2008, Rix2013}.  

This classical picture of the Milky Way reveals a rather complex galaxy, with four primary ``internal'' components -- the bulge, the thin disk, the thick disk and the halo, and one ``external'' component -- the accreted material. These components form the basis of any attempt to model, to simulate and to understand the formation and evolution of our Galaxy \citep[see e.g. the reviews of ][and references therein]{Freeman2002, Ivezic2012}. These components are usually defined by the phase spaces in which we observe them (spatially, kinematically, and chemically). However there is an ongoing debate as to whether these components are truly distinct objects or a part of a continuous sequence of an evolving galaxy \citep[e.g.][]{Bovy2012, Rix2013}. 

Among the phase spaces to `separate' the Galactic components, the method of using chemical abundance ratios as a means to probe the Galactic components has been performed and advanced significantly in the past decade by using relatively large samples of high resolution spectra \citep[e.g.][]{Edvardsson1993, Fuhrmann1998, Venn2004, Nissen2010, Sheffield2012, Ramirez2012, Feltzing2013, Bensby2014, Mikolaitis2014}. In particular, the ratio of the mean of the $\alpha$-elements\footnote{e.g. Ca, Mg, Ti, O, S, and Si which are thought to be produced via Type II supernovae (SNII)} relative to iron has been shown to be very powerful to disentangle the various Milky Way components. Active research has been dedicated in order to explain the two sequences found in the abundances of the $\alpha$-elements respect to the abundances of iron. On one hand, at low metallicities ([Fe/H]~$\sim-1.2$ dex), the low--$\alpha$ sequence is attributed to stars that were chemically enriched in an environment separate from that of the bulk of the halo and later accreted on to the Galactic halo. At higher metallicities ([Fe/H] $\sim -0.5$ dex) the \apoor stars are interpreted as the thin disk. On the other hand, at low metallicities the high-$\alpha$ sequence is attributed to the halo and at high metallicities the high-$\alpha$ sequence is attributed to the thick disk although there is likely significant overlap between the two.

\subsection{The canonical picture of the Galactic components}
\label{subsec:introcomponents}
Classically, the Galactic disk is often split into a thin disk and thick disk component \citep[e.g.][]{Gilmore1983,Freeman2002, Rix2013}. The thin disk is thought to contain mostly young stars and is likely the final stage of dissipative collapse. It is defined spatially by an exponential power law with a small vertical scale height and large radial scale length \citep[e.g.][]{Bovy2012b, Haywood2013}. Kinematically the thin disk stars follow near circular, co-rotational orbits with a low velocity dispersion \citep[e.g.][]{Edvardsson1993, Reddy2003, Kordopatis2013b, Rix2013}. Chemically, the thin disk its thought to extend over a metallicity range of +0.1 $<$ \feh\ $<$ --0.70 dex and is near solar values in the $\alpha$-elements \citep[e.g.][]{Bensby2014}. 

The thick disk was initially found to be separate from the thin disk on the basis of the spatial distribution of its stars \citep[e.g.][]{Yoshii1982, Gilmore1983}. It is thought to form potentially through a variety of processes including: satellite heating, accretion, or merging induced star formation, secular disk heating, radial migration, and others \citep[for a review of these mechanism we refer the reader to][and references therein]{Rix2013}. However the importance of each of these mechanisms in the formation and assembly of the thick disk is still not well understood. It is defined spatially by an exponential power law with a larger vertical scale height and smaller radial scale length compared to the thin disk \citep[e.g.][]{Bovy2012b, Haywood2013}. Kinematically, thick disk stars co-rotate with the disk, albeit with a smaller rotational velocity and overall have hotter orbits than their thin disk counterparts \citep[e.g.][]{Edvardsson1993, Reddy2003, Bensby2005, Haywood2013, Kordopatis2013b}. Chemically, it is thought to be distinct from the thin disk component in the \afenospace-metallicity plane with a high \afe signature \citep[e.g.][]{Bensby2014, Nidever2014, Recio-Blanco2014}. The thick disk has a metallically that extends significantly lower than the thin disk, and can extend down to well below [Fe/H] $<$ --1.0 dex \citep[e.g.][]{Beers2002}. The thick disk is also thought to be older than the thin disk \citep[e.g.][]{Haywood2013, Masseron2015}. 

The Galactic halo is thought to naturally form very early on from a mixture of dissipative collapse and accretion \citep[e.g.][]{Eggen1962, Searle1978, Ibata1994, Belokurov2006}. As such most of its stars are very old \citep{Jofre2011, Hawkins2014}. It is spatially defined by a power-law with an index of approximately --~3.5 \citep[e.g.][]{Helmi2008}. Kinematically, the (inner) Galactic halo is thought to be pressure supported, have a small net prograde rotation \citep[e.g.][]{Carollo2010}, and contain the highest velocity stars \citep[e.g.][]{Hawkins2015}. Chemically, the Galactic halo is predominantly metal-poor \citep[e.g.][]{Schlesinger2012} and enriched in the $\alpha$-elements \citep[e.g.][]{Ishigaki2012}. The (inner) Galactic halo has been shown to split into two chemically distinct components. Several recent studies have reported the presence of \apoor stars in halo samples \citep{Nissen2010, Schuster2012, Ramirez2012, Sheffield2012, Bensby2014, JacksonJones2014, Hawkins2014}. These studies have shown that the \apoor sequence is distinct in kinematics, ages, and other chemical elements such as C, Na, and Ni compared to the \arich sequence. It is thought that the \apoor sequence is assembled through the accretion of satellite galaxies.

\subsection{Decomposition of the canonical Galactic components}
\label{subsec:decomp_intro}
Decomposing these components chemo-kinematically, particularly in the region of metallicity space where they overlap (--1.2 $<$ \feh\ $<$ --0.60 dex) is a significant challenge yet critical to understand the formation and assembly of the Galaxy, as this is the region of metallicity space where the thin disk, thick disk, and halo are all assembling and co-evolving. The problem arises with the fact that the original dynamical and spatial distributions of stars are perturbed over time, potentially erasing the memory of the original sites. Kinematical heating and spatial disruption can be produced via several process, such as bar resonances and radial mixing, clumpiness in the gas caused, and minor mergers, to name a few \citep[e.g.][]{Minchev2012, Rix2013, Haywood2013, Sanders2013, Sanders2015}.  Furthermore, dynamical and spatial distributions depend heavily on the distance of the stars, which is subject of large uncertainties for the majority of them \citep[e.g.][and references therein]{Binney2013review}. Thus, matching models and simulations to observational data continues to be one of the fundamental challenges in Galactic astronomy today. 

\begin{figure*}     
	\includegraphics[scale=0.35]{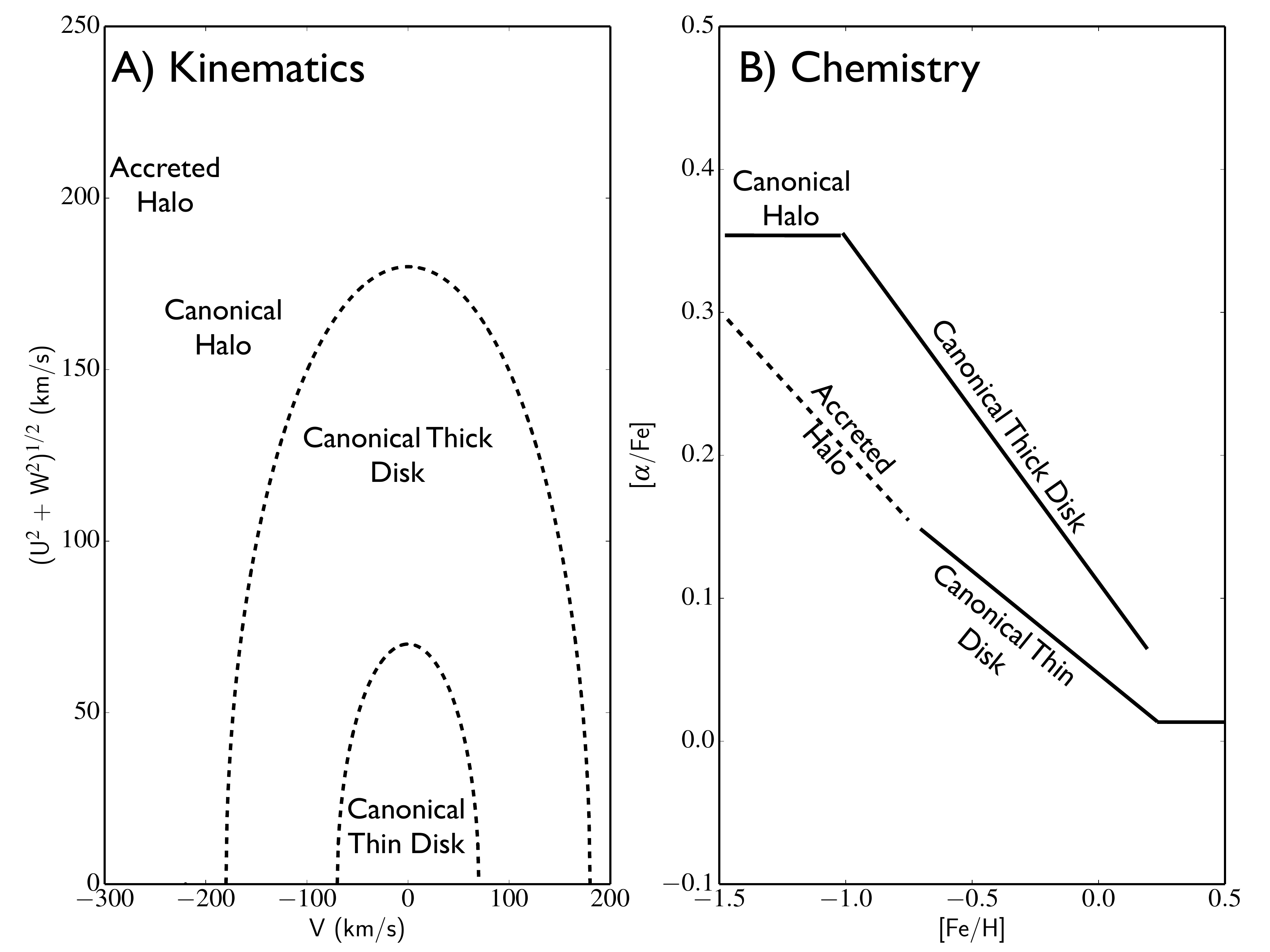}
  	\caption{A) A cartoon illustration of the decomposition of the Galactic components along the Toomre diagram which plots the quadrature of the radial (U) and vertical (W) velocities as a function of the velocity along rotation (V). All velocities in the Toomre diagram are relative to the LSR. (B) A cartoon illustration of the decomposition of the Galactic components along the \afenospace-metallicity plane. } 
	\label{fig:decomp_cartoon}
	 \end{figure*}

Despite this, the current mode through which Galactic components are decomposed is through kinematics via the Toomre diagram with aid of the \afenospace-\feh\ plane. The left panel of Figure \ref{fig:decomp_cartoon} shows an illustration of the Toomre diagram which plots the quadrature of the vertical (denoted by W) and radial (U) velocities as a function of the velocity along Galactic rotation (V). In this case, these velocities are relative to the local standard of rest (LSR). This diagram or a similar probabilistic kinematic Galactic component decomposition is widely used in the literature \citep[e.g.][]{Bensby2003, Venn2004, Nissen2010, Schuster2012, Ishigaki2012, Bensby2014, Hawkins2015}. The power of the Toomre diagram lies in the fact that the canonical thin disk, thick disk, and halo have increasingly hotter kinematics (i.e. each component lies in an increasingly larger constant velocity circle on the Toomre diagram) and thus are further elevated in the diagram. The fundamental drawback to this diagram is the need for accurate proper motion and distances to fully resolve the 3D velocity vector. 

The \afenospace-meallicity plane (a cartoon version of this is shown in the right panel of Figure \ref{fig:decomp_cartoon}) is another approach to decompose the Galactic components. This method works because the only ``easily measurable'' property of long-lived stars that remains relatively unchanged over cosmic times is their chemical patterns that are imprinted in their spectra. Current large stellar spectroscopic surveys of the Milky Way, especially those of high resolution like APOGEE \citep{Eisenstein2011}, GALAH \citep{De_silva2015} or Gaia-ESO \citep{Gilmore2012}, will enable methods such as chemical tagging stars providing us with a new picture for our Galaxy \citep[e.g.][]{Feltzing2013}. Under this scheme, a given stellar population is marked by the chemical distribution of the medium from which the stars formed. This distribution is principally defined by the initial mass function of the previous generation of stars that enriched the gas, the star formation rate and the different yields from the nuclear reactions. 
\\ \\
In this work, we study the stars at the Galactic disk(s)-halo interface by taking advantage of the APOGEE public spectroscopic survey \citep{Eisenstein2011}, which contains not only thousands of stars at this interface, but its very high-resolution spectra allow us to go far beyond the study of only $\alpha$-elements. Specifically we are interested to know if the Galactic components must be defined in chemo-kinematic-spatial spaces or if they can be defined purely by chemistry. In this paper, we present a detailed chemical abundance evolution analysis of 14 different elements available in the APOGEE survey (including $\alpha$, odd-even, light, and Fe-peak elements) with the aim to discuss how the thin disk, the thick disk, the halo and the accreted halo stars are linked. In particular, we are interested in the $\alpha$-elements, carbon, nickel (and other Fe-peak elements), and other odd-even elements as they can be used to distinguish the accreted halo from the canonical halo, and canonical thin disk from the canonical thick disk \citep[e.g.][]{Bensby2014}. In section \ref{sec:DATA} we describe APOGEE data selection criteria and the subsample definitions. In section \ref{sec:chemistry}, we compare and contrast, element-by-element, the various Galactic components in order to better understand how they are linked. We then use the element-by-element analysis in order to discuss our results compared to the current literature on the halo-disk transition regions in section \ref{sec:discussion}. In section \ref{sec:chemtag} we use the abundance ratio trends to propose a new powerful chemical labeling technique that uses only chemistry to efficiently separate the various Galactic components. Finally, we summarize our findings in section \ref{sec:summary}.

\section{Data and Subsamples}
\label{sec:DATA}
\subsection{Data: The APOGEE Survey}
\label{subsec:data}

To study the interface of the Galactic disks and halo, we employed the first two years of data from the SDSS-III/APOGEE survey \citep[described in][]{Eisenstein2011}. APOGEE used $H$ band spectroscopy to determine the stellar parameters, chemical abundances, and radial velocities for a large sample of stars. The most current data release (SDSS DR12) contains nearly 100 000 stars with stellar parameters, chemical abundance determinations and spectra \citep{Holtzman2015}. The typical uncertainties in the parameters are $\pm$100~K, $\pm$0.1~dex, $\pm$0.04~dex, and $\pm$0.04~dex in \Teffnospace, log g, [Fe/H], and \afenospace, respectively. While there may be some systematics biases 
\citep[potentially up to +0.20 dex, see][]{Masseron2015}, the precision of APOGEE chemical abundances are expected to be relatively high (with typical internal uncertainties of less than 0.10 dex). To minimize the potential uncertainties in the accuracy in the abundance scales, in this paper we have done a relative abundance analysis of the various Galactic components. 

We selected a sample of giant stars, beginning with the full APOGEE sample and using the following cuts:
\begin{itemize}
\item  The signal-to-noise ratio (SNR) had to be larger than 100. We used this cut to ensure that the selected stars have quality estimates of the stellar parameters and \afenospace. 

\item \Teff $>$ 4000 K. This cut was chosen because as noted by \cite{Holtzman2015}, stars with \Teff $<$ 4000 K will likely have larger uncertainties in stellar parameters and chemical abundance. 

\item The ASPCAP, metallicity, and \afe flags must be set to 0. This cut ensured that there were no major flagged issues (e.g. low SNR, poor synthetic spectral fit, stellar parameters near grid boundaries, etc.).

\item 1.0 $<$ log g $<$ 3.5 dex. This cut was used to deselect any dwarf stars for which the stellar parameters and chemical abundances would likely have large uncertainties \citep[e.g.][]{Nidever2014, Holtzman2015}.

\item The stars had to be located at $b<60$ degrees. This was needed to separate disk-like from non-disk like kinematics in $l$-Galactic rest frame radial velocity (GRV) space as we discuss in section \ref{subsec:kinematics}. 
\end{itemize}

The initial sample contained about 69000 stars. The sample was then restricted to the metallicity regions --1.20 $<$ [Fe/H] $<$ --0.55 dex in which encompasses the transition between the Galactic disks and halo. This reduced the sample to about 3200 stars. We note that we used the uncalibrated [Fe/H] for metallicity rather than the calibrated global metallicity, [M/H], for the purposes of comparing [X/Fe] abundance ratios to literature. \cite{Holtzman2015} noted that the APOGEE uncalibrated [Fe/H] tends to be overestimated at low metallicities. However, applying the calibration to [Fe/H] would cause the [X/Fe] ratios to be inflated compared to literature values. In this paper we focus on a relative abundance analysis within subgroups, and as such the use of [Fe/H] as opposed to [M/H] only changes the absolute abundance scale but does not change the relative analysis leaving our conclusions robust. See section \ref{subsec:thindisk} for more details. In the following section we use the chemistry and kinematics to decompose the sample into the various canonical Galactic components (similar to the scheme outlined in Figure \ref{fig:decomp_cartoon}).  

\subsection{Chemokinematic decomposition of canonical Galactic components}
\label{subsec:kinematics}
\begin{figure}
 	\centering
	 \includegraphics[width=\columnwidth]{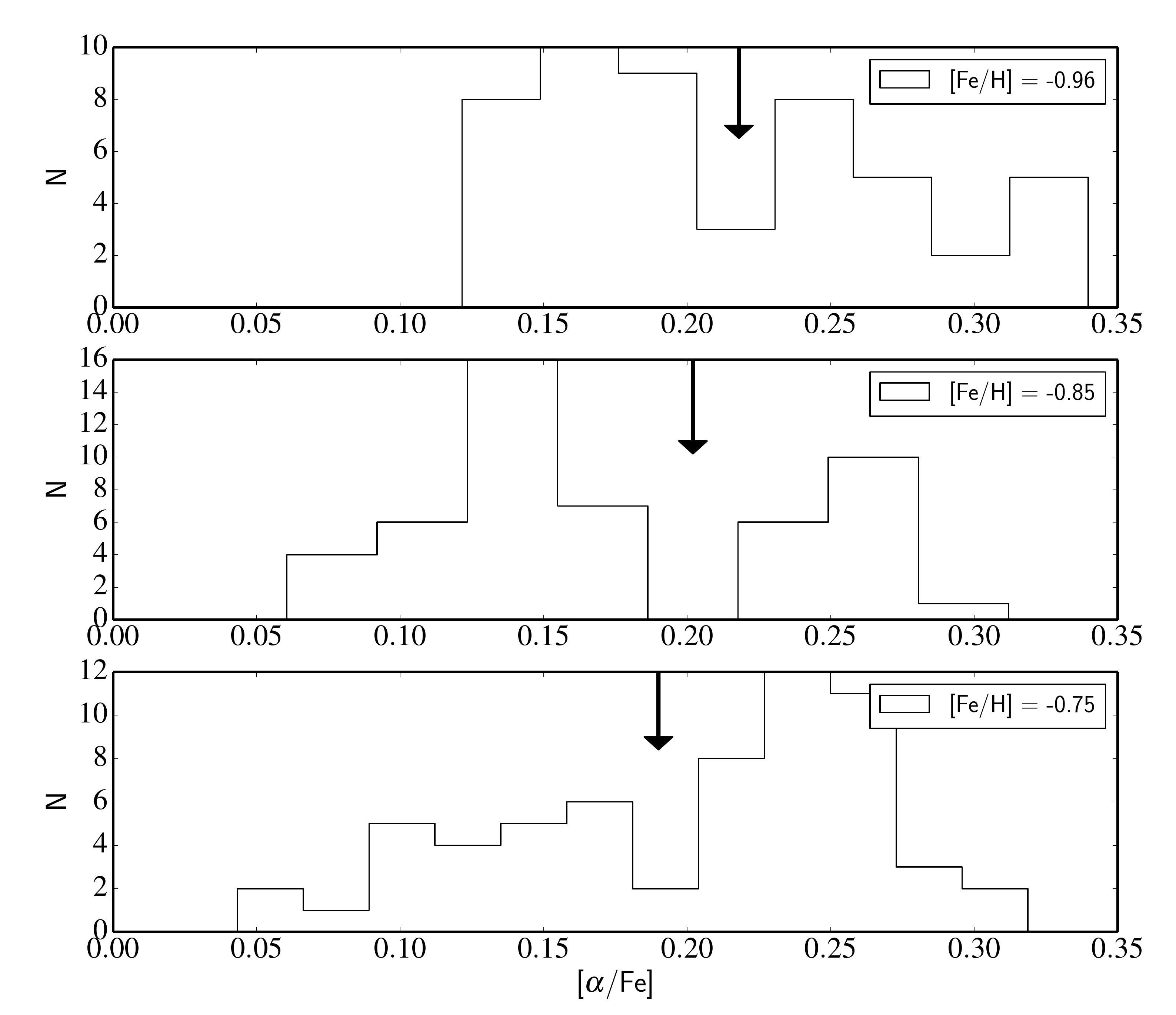}
	\caption{The distribution of \afe\ in three metallicity bins, with bin centers at \feh\ = $-0.96, -0.85,-0.76$ dex respectively. In each metallicity bin there appears to be a bimodal distribution in \afe\ the trough between the \apoor and \arich components are marked by a black arrow.}
	\label{fig:alphadist_binned}
\end{figure}

\begin{figure*}     
       \subfigure[]{
              \includegraphics[width=2\columnwidth]{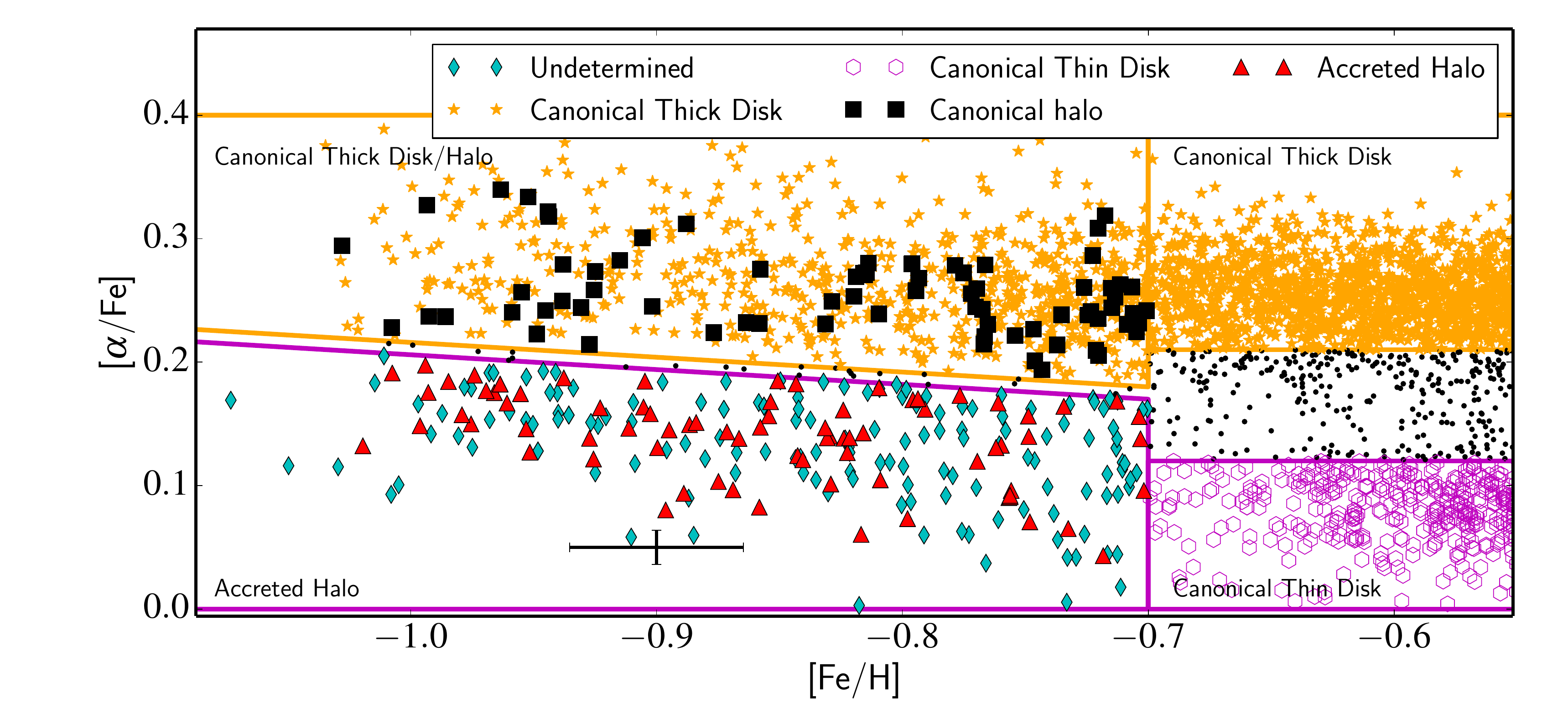}
                \label{fig:alphamet}  }
        \subfigure[]{
              \includegraphics[width=2\columnwidth]{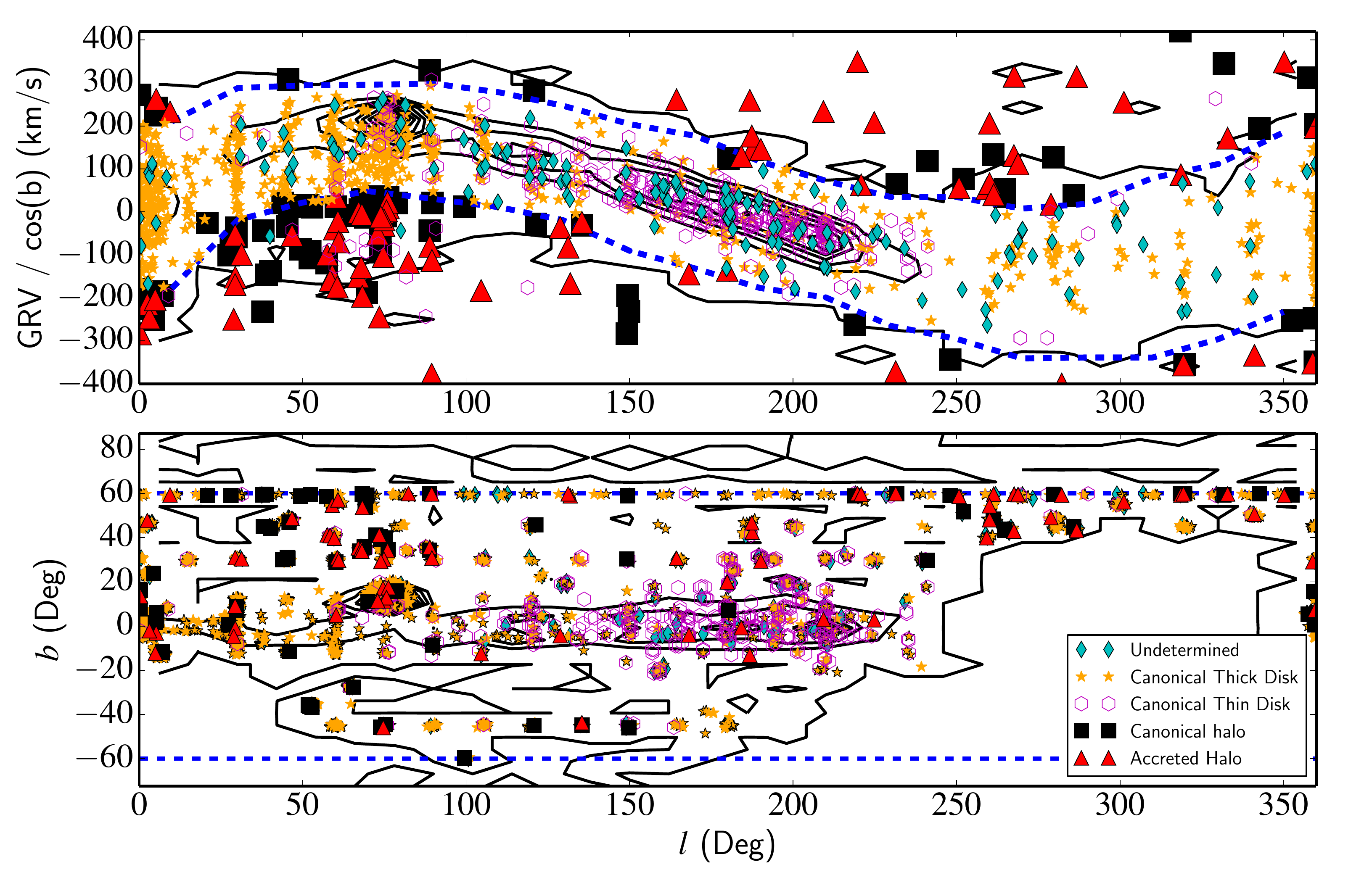}
               \label{fig:lGRV}   }

        \caption{(a) The \afe as a function of metallicity for the intermediate metallicity  APOGEE sample. We separated chemical the various Galactic components based on where they sit in \afenospace-metallicity diagram. A cut is placed at \feh\ = --0.70 dex as this where the thin disk is thought to end \citep[e.g.][]{Bensby2014}. The error bar represents the median uncertainty in \afe and metallicity. The black points are the stars that were deselected because it is difficult to classify those stars on the basis of their chemistry. (b) Upper panel: The GRV as a function of $l$ for the full APOGEE sample that pass our quality cuts (contours). The dotted blue lines represents 2.5 $\times$ disk GRV dispersion in each $l$ bin. It is clear from this diagram that the disk signature is a sinusoidal pattern in $l$-GRV space. Therefore we select stars that have absolute GRVs larger than dotted blue line as our non-disk subsamples (i.e. these stars have, on average less than a 1.24 per cent chance of being kinematically a part of the disk), which are marked as black squares and red triangle for the \apoor and $\alpha$-rich, respectively. While the orange stars and cyan diamonds are the \arich and \apoor stars that have disk-like kinematics (i.e. they are inside of the blue dotted line) and thus are consistent with disk-like kinematics. Stars with metallicities larger than --0.70 dex that are \apoor are considered canonical thin disk stars (open magenta hexagons) Lower Panel: the $l$, $b$ Galactic coordinates of the full APOGEE sample that pass our quality cuts (contours), following the same symbols as the top panel.}
        \label{fig:alphamet_lGRV}
\end{figure*}

Since we are primarily interested in the disks-halo interface, our focus in this paper is on the intermediate metallicity (--1.20~$<$ \feh\ $<$~--0.55 dex) region. In Figure \ref{fig:alphamet} we plot the standard \afenospace-metallicity for the intermediate metallicity stars. With our initial sample we recovered the two disk sequences discussed in section \ref{subsec:decomp_intro} and seen in other APOGEE studies \citep[e.g][]{Nidever2014, Masseron2015}. In the low-metallicity regime (--1.20~$<$ \feh\ $<$~--0.70 dex), we found the two sequences: an \apoor sequence (left magenta selection box) and an \arich sequence (the left orange selection box) similar to \cite{Nissen2010}. We determined the boundary between these two sequences by identifying the trough in the \afe distribution in three metallicity bins and performing a linear fit in the \afenospace-metallicity plane of these three metallicity bins similar to \cite{Recio-Blanco2014}. In Figure \ref{fig:alphadist_binned}, we show the \afe\ distribution in each of the metallicity bins used to determine the separation between the \apoor and \arich sequences. The bins were chosen to ensure at least 50 stars were in each bin. We note that the bimodality is most clear when all halo stars (i.e. those stars assigned only to canonical and accreted halos not including stars from the disks and undetermined groups) are used without binning in metallicity space.

With the lack distances and proper motions we do not employ the Toomre diagram as a means assign the stars to a component because we cannot estimate their full 3D velocity. However, we made use the RVs, which have typical uncertainties on the order of 0.1 \kmsnospace. We used a method that combines the $l$ and GRV in order to disentangle the halo from disk stars similar to \cite{Sheffield2012}. We computed the GRV by correcting the radial velocity for solar motion \cite[see Equation 1 of][]{Hawkins2015}. For this, we assumed $\vec{v}_{\sun} = [U_{\sun}, V_{\sun}, W_{\sun}] = [14.0, 12.24, 7.25]$ km~s$^{-1}$, V$_{\mathrm{LSR}}$ = 220 \kms \citep{Schonrich2012}. In Figure \ref{fig:lGRV} we show the GRV/cos($b$) as a function of $l$ for the full APOGEE sample that pass our quality cuts (black contours). As noted by \cite{Majewski2012} and \cite{Sheffield2012}, dividing by the GRV by the cos($b$) accentuates the difference between the disk and non-disk components provided the data is restricted to |$b$| $< 60$ degrees allowing us to separate out halo stars more efficiently. This cut, while potentially eliminating some halo stars, still leaves enough to study the chemical abundance trends. We have also tested several |$b$| cuts between 40 -- 60 deg with little to no affect on the chemical abundance trends and overall conclusions of this work. As |$b$| approaches larger values, the cos($b$) becomes too small relative to the GRV to be useful.

In $l$-GRV space (Figure \ref{fig:lGRV}), the Galactic disk makes a sinusoidal pattern with an amplitude approximately equal to the rotational velocity of the Galaxy and dispersion correlated to the velocity dispersion of the disk. In the same space the Galactic halo has a more-or-less random distribution \citep[e.g.][]{Johnston2012}. To characterize the disk in this space, we selected a subsample of APOGEE stars with metallicities between --0.50 and 0.00 dex (i.e. the metallicity range where the disk dominates) and binned in $l$ space (20 degrees bins). The dotted blue lines in the top panel of Figure \ref{fig:lGRV} are equal to 2.5 times the GRV dispersion in each $l$ bin which represents stars with disk-like kinematics. Stars that have an absolute GRV/cos($b$) that is at least 2.5$\sigma$ or larger than the Galactic disk within a given $l$ bin are considered to have non-disk like kinematics and are assigned to the halo. We note that there are a handful of stars which are assigned to the Galactic halo which appear to sit below GRV dispersion cut in Figure \ref{fig:lGRV}. These stars are on the $l$ bin edges and assigning them to the halo or disk does not impact the results given the large sample size. While there still may be some disk contamination in the halo sample, this is likely to be small. Similarly, there may be halo stars with disk-like kinematics that this study also rejected. Full 3D space motions could aid in recovering these lost halo stars while further rejecting thick disk stars. We varied the GRV dispersion cut from 1 -- 4 $\sigma$ to asses the impact of the kinematic classification. We discuss the impact of the kinematic selection in section \ref{sec:discussion}, however we note here it does not affect our conclusions.

As we are interested in studying the bulk properties of the thin-thick disk-halo transition, we must use the chemodynamical information to define our sample into the canonical Galactic components discussed in sections \ref{subsec:introcomponents} and \ref{subsec:decomp_intro}. Our selection is inspired by the literature described in section \ref{subsec:introcomponents}, however it is important to keep in mind that our definitions may not be the same as other studies thus comparisons should be done with care. We define the Galactic components in our sample in the following way:
\begin{itemize}
\item \textbf{Canonical Thin Disk (open magenta hexagon)} -- These are stars (shown as the right magenta selection box in Figure \ref{fig:alphamet}) that have low \afe ratios \citep[\afe $<$ +0.10 dex, e.g.][]{Adibekyan2012, Haywood2013, Recio-Blanco2014} and have metallicities $>$ --0.70 dex \citep[e.g.][]{Bensby2014}. The cut in \afe was designed to conservatively select out the Galactic thin disk. Kinematically these stars have spatial-kinematic coherence (falling inside of the blue dotted line in Figure \ref{fig:lGRV}). We note that a result of cutting our sample at \feh\ $<$ --0.55 dex and the negative radial metallicity gradient found in the disk \citep[e.g.][]{Cheng2012}, the thin disk stars in our sample are preferentially found in the outer Galaxy.

\item \textbf{Canonical Thick Disk (orange stars)} -- These stars (orange selection boxes in Figure \ref{fig:alphamet}) have high \afe ratios \citep[\afe larger than +0.22 dex at \feh\ $>$ --0.70 dex, e.g.][]{Adibekyan2012, Haywood2013, Recio-Blanco2014}. Kinematically these stars have spatial-kinematic coherence (falling inside of the blue dotted line in Figure \ref{fig:lGRV}). 

\item \textbf{Canonical Halo (black squares)} -- These stars are defined as \arichnospace, metal-poor stars (left orange selection box in Figure \ref{fig:alphamet}). The \afe cut was determined by a linear fit to the trough in the \afe distribution in three metallicity bins below --0.70 dex, where the halo and thick disk will dominate. Kinematically they were selected to have high velocities \citep[e.g.][]{Nissen2010, Schuster2012, Hawkins2015} and not have significant spatial-kinematic coherence (falling outside of the blue dotted line in Figure \ref{fig:lGRV}). Additionally, there may be some contamination of this component from the accreted halo caused by the `hard' cut used to separate the two. However, less than a fifth of the canonical halo stars fall within 1-$\sigma$ of the \afe\ boundary. 

\item \textbf{Accreted Halo (red triangles)}-- These stars are \apoor and metal-poor (left magenta selection box in Figure \ref{fig:alphamet}). They were selected kinematically to have high velocities \citep[e.g.][]{Nissen2010, Schuster2012} and not have significant have spatial-kinematic coherence (falling outside of the blue dotted line in Figure \ref{fig:lGRV}). Our choice to use a `hard' boundary to separate the \apoor accreted halo and the \arich canonical halo \citep[as done in other literature, e.g.][]{Nissen2010, Nissen2011, Schuster2012, Bensby2014} may result in some contamination of the accreted halo subsample from the \arich canonical halo (or thick disk in some cases) or vice-versa. Accreted halo stars within $\sim$ 0.02 dex of the boundary, the typical 1-$\sigma$ uncertainties in \afenospace, could potentially be true canonical halo stars. This only accounts for less than a quarter of the total accreted halo sample. Thus the canonical halo is not likely to be a large contaminating source in the accreted halo.

\item \textbf{Undetermined (cyan diamonds)}-- These stars are \apoor (right magenta selection box in Figure \ref{fig:alphamet}) with metallicities less than --0.70~dex but may have spatial-kinematic coherence (falling inside the blue dotted line in Figure \ref{fig:lGRV}). In this way these stars have the chemistry of what would be the accreted halo but kinematics that cannot rule out disk-like motion. The lowest metallicity thin disk stars have been found to have \feh\ = --0.70~dex \citep[e.g.][]{Bensby2014}, thus these undetermined stars could be a part of the accreted halo or misidentified thick disk stars with lower \afenospace. Additionally the undetermined stars within $\sim 0.02$ dex of the boundary separating \arich and \apoor stars may be canonical halo stars however, the lower the \afe of the undetermined stars the less likely that this is the case.
\end{itemize}

The halo is thought random in $l$-GRV space \citep[e.g.][]{Johnston2012} and thus with a lack of full 3D space motions there is a degeneracy that could result in some contamination of halo stars in all canonical disk subgroups. However, even though the halo is thought to extend to metallicities as large as --0.50 dex \citep[e.g.][]{Nissen2011}, the Galactic disk populations will dominate at metallicities above --0.80 dex \citep[e.g.][]{Schlesinger2012} particularly if they follow the sinusoidal disk-like pattern in Figure \ref{fig:lGRV}. A rough estimate of the upper limit of the contamination of the disk components on the halo components as a result of using the $l$-GRV diagram to decompose the Galaxy can be determined using the procedure outlined in section 5 of \cite{Sheffield2012}. Assuming that all 3200 stars in our sample are disk stars, and that the stars in a given $l$ bin are distributed in a Gaussian way around the underlying sinusoidal `disk-like' feature, $\sim$ 40 of those stars (i.e. 1.2 per cent or 2.5 $\sigma$) would be identified as `halo' stars. However, we found nearly 160 potential halo stars of which we estimate at most 40 of these could be true disk members. This would yield a contamination rate of 25 per cent. Additionally, we used the Galaxia Galactic model \citep{Sharma2011} to further study the expected contamination rate. The models have been set to have a metallicity distribution for each component described in section 3.1 of \cite{Kordopatis2013b}. We used the same magnitude and color cuts of APOGEE. We applied the same selection routine as above. Results indicate that the contamination rate of the (thick) disk on the Galactic halo samples can range between 20--60 percent depending on the maximum distance surveyed. Larger sampling distance yield lower contamination rates. We also note that adopting a 3-$\sigma$ cut, instead of 2.5-$\sigma$, reduces this contamination to in most cases below 15 per cent. As such, we varied the GRV dispersion cut to as high as 4-$\sigma$ (changing the amount of contamination in the disk and halo subgroups) and found that the chemical abundance trend and the overall conclusions are not significantly affected.

Our final sample contains a total of approximately 150 undetermined stars (cyan diamonds) in the low-$\alpha$, low-metallicity region with disk-like kinematics that could be thin disk stars or accreted stars. There are also approximately 160 halo-like stars split in almost equal parts between the accreted halo (red triangles) and the canonical halo components (black squares). The accreted halo seem to be distributed over all $l$, $b$ space that has been sampled. Unsurprisingly there are more halo stars at high Galactic latitudes. Accurate distance and proper motions could help compare the total space motions and orbital parameters of the various components, which is beyond the scope of this paper.  There are approximately 370 canonical thin disk stars (open magenta hexagons), and 1400 canonical thick disk stars (orange stars).  

  \section{Analysis of Chemical Abundances of the Canonical Galactic Components}
\label{sec:chemistry}
In this section we discuss the individual abundance ratios and trends as a function of metallicity of the five identified subsamples (i.e. canonical thin disk, canonical thick disk, halo and accreted halo and undetermined) defined in section \ref{subsec:kinematics}. Although one may be concerned by the systematic errors of the absolute abundances in the APOGEE data, our study mainly relies on relative comparison between the different populations. Therefore, our discussion and conclusions should not be affected by these systematic errors; in our case only the relative abundance precision matters.

\subsection{The $\alpha$-elements: O, Mg, Si, S, Ca, and Ti}
\label{subsec:alphael}
We study the $\alpha$-elements as a way to infer the properties of the size and star formation rate of the cloud that formed the different stellar populations. The $\alpha$-elements are those elements that are, by definition, made by capturing $\alpha$-particles. This implies that those elements have an atomic mass number that is a multiple of 4, thus $\rm ^{16}O, ^{24}Mg, ^{28}Si, ^{32}S, ^{40}Ca, and \ ^{44}Ti$. While other isotopes of those elements exist, they cannot currently be distinguished in the APOGEE dataset, as such we will assume that the listed isotopes are the dominant contributor to the elemental abundances of the APOGEE data. 

Those elements are made in the cores of stars by $\alpha$-capture during post-main sequence burning and dispersed in the interstellar medium via type II supernovae (SNII), while relatively less is expected to be produced by type Ia supernovae (SNIa). Thus the ratio of \afe is thought to be sensitive to the environment enrichment history of the gas from which the star formed. Unlike the \apoor stars, the \arich stars are thought to be formed in regions where SNII were numerous, thus in a significantly higher star formation than the \apoor stars \citep[e.g.][]{Gilmore1998}.

\begin{figure*}
	 \includegraphics[width=2\columnwidth, height=0.9\textheight, keepaspectratio]{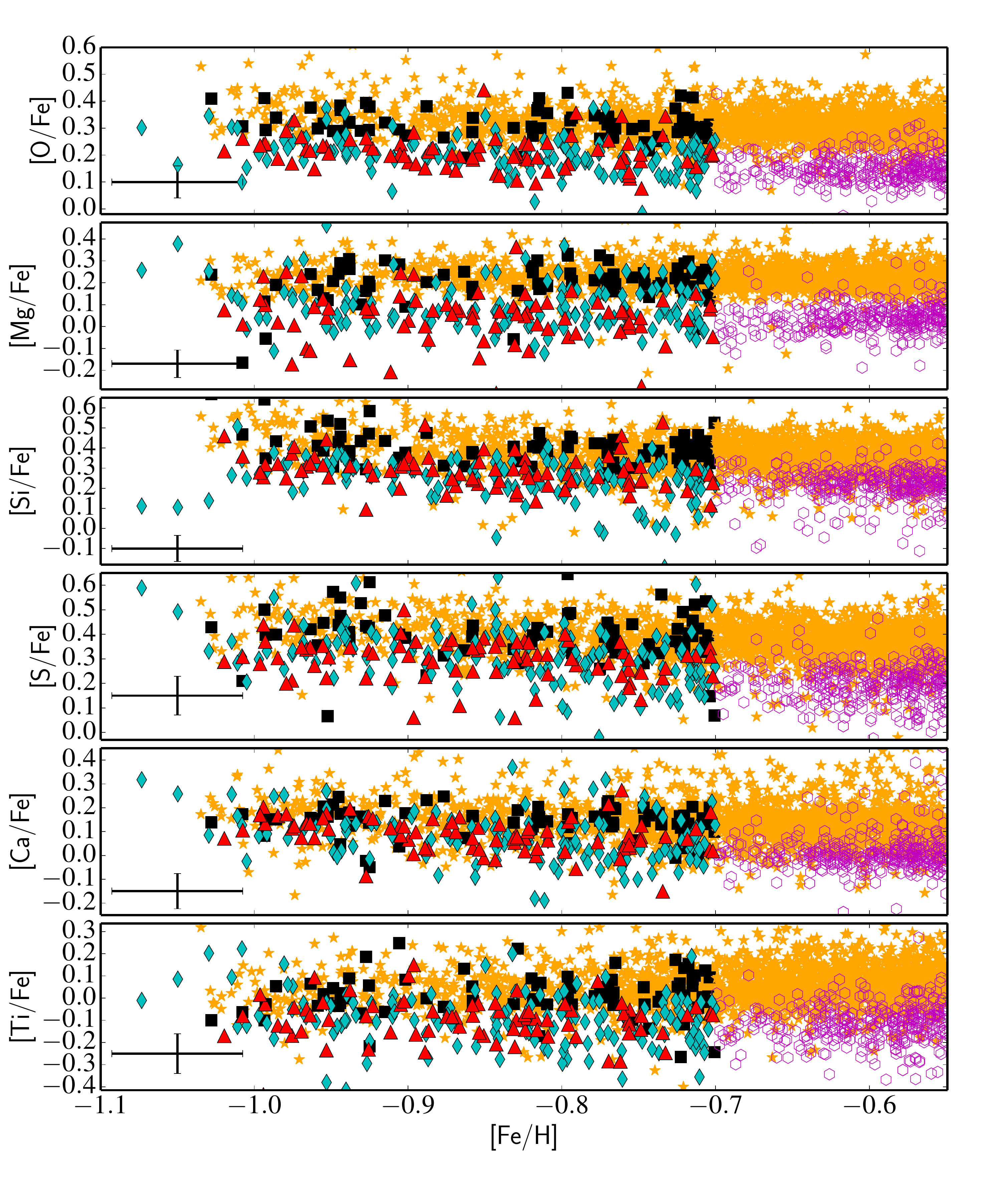}
	\caption{The [O/Fe], [Mg/Fe], [Si/Fe], [S/Fe], [Ca/Fe], [Ti/Fe] from top to bottom respectively as a function of metallicity for the accreted halo (red triangles), canonical halo (black squares), canonical thick disk (orange stars) and undetermined (cyan diamonds), and canonical thin disk (magenta open hexagons) subgroups. It is clear that the distinction between the \arich and \apoor sequence at low metallicity is likely driven by the [Mg/Fe] and [O/Fe] abundance ratios. }
	\label{fig:alphachem}
\end{figure*}

In Figure \ref{fig:alphamet} we see that indeed APOGEE data shows evidence of an \apoor and \arich sequence at both the intermediate and low metallicity regimes, indicating a distinct formation history for each sequence \citep{Nidever2011}. In Figure \ref{fig:alphachem} we plot the abundance of each $\alpha$-element relative to iron as a function of metallicity in a different panel, ordered by atomic number. The different symbols represent the populations defined in section~\ref{subsec:kinematics}. The abrupt change of symbols at [Fe/H] = --0.70 dex is due to the definition of the populations only. 

Unlike several Galactic studies in the literature \citep[e.g.][]{Reddy2003, Venn2004, Haywood2013, Bensby2014}, our sample contains the $\alpha$-element sulphur. Not surprisingly, it behaves in a similar way to the other $\alpha$-elements \citep[e.g.][]{Nissen2007, Spite2011, Jonsson2011, Takada-Hidai2012, Matrozis2013, Skuladottir2015} yet not used in the standard \afe definition. Since we used this the global \afe criteria to define our populations, we obviously find that our populations separate in each of the individual $\alpha$-elements.

 From Figure~\ref{fig:alphachem} we also note that in the --0.90 $<$ \feh\ $<$ --0.70 dex metallicity regime, O and Mg drive the largest distinction between the \arich and \apoor sequences. The rest of the $\alpha$ element abundances appear to be very mixed, making it difficult to disentangle a clear $\alpha$-poor sequence from an $\alpha$-rich one. This could be a result of O and Mg SNII yields being mass-dependent \citep[e.g.][]{Nomoto2013}.

\subsection{The Fe-peak elements:  Mn, Ni}
\label{subsec:Fe-peak}
\begin{figure*}
\includegraphics[width=2\columnwidth]{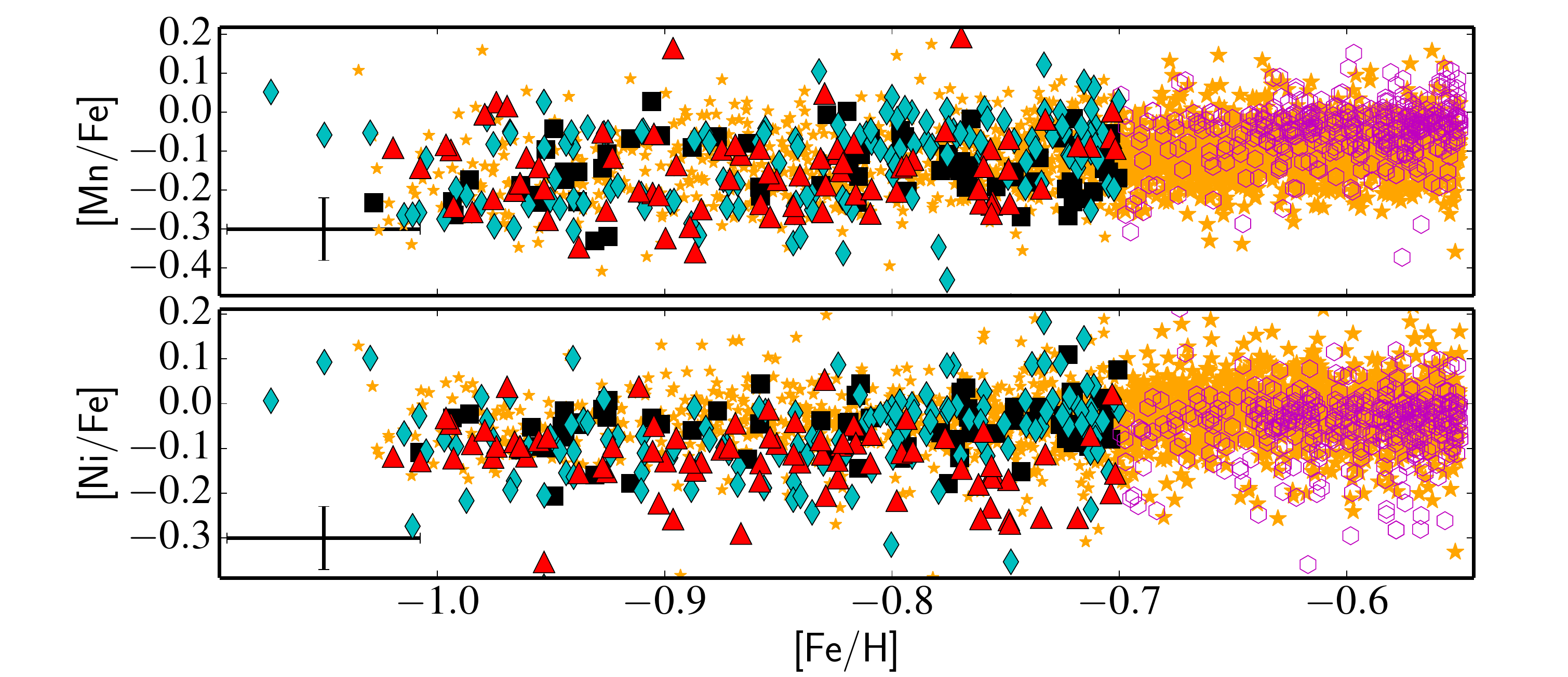}
\caption{The [Mn/Fe], [Ni/Fe] from top to bottom, respectively as a function of metallicity for the five groups with the same symbols as Figure~\ref{fig:alphachem}.}
\label{fig:Fe-peak}
\end{figure*}

In contrast to the $\alpha$-elements, Fe-peak elements, such as Mn and Ni, are thought to be formed primarily via SNIa \citep[e.g.][]{Iwamoto1999}. Therefore they are expected to track Fe in such a way that [Mn/Fe] has been shown to be sub solar in low metallicity stars and increase as with [Fe/H] \citep[e.g.][]{Adibekyan2012, Battinstini2015}.  We plot in Figure \ref{fig:Fe-peak} the abundances of the only Fe-peak elements in APOGEE, namely Mn and Ni, for the intermediate metallicity stars as a function of [Fe/H].  As in Figure~\ref{fig:alphachem}, the different symbols represent our definition of populations (see Sect.~\ref{subsec:kinematics}). 

It is interesting to note that Mn has a different behaviour for thin and thick disk stars (orange and magenta colour, respectively) at comparable metallicities. While thin disk stars have Mn enhanced respect to iron, thick disk stars have it at the same level than iron. Since Mn is produced at a higher fraction compared to Fe during SNIa \citep[e.g.][]{Gratton1989, Iwamoto1999, Kobayashi2006} one may expect that at a given metallicity, \apoor stars -- which have been polluted by more SNIa -- will have higher [Mn/Fe] ratios compared to their \arich counterparts. This offers an explanation as to why the canonical thin disk stars seem to have overall higher [Mn/Fe] ratios compared to the canonical thick disk. \citet{Kobayashi2011} demonstrate from models that O and Mn should be anti correlated in the Galaxy, because O is produced mainly by SNII while Mn is produced predominantly by SNIa. We observe in Figure \ref{fig:alphachem} that the canonical thin disk has lower [O/Fe] than its thick disk counterpart. Thus, given the results of \cite{Kobayashi2011}, one would expect the thin disk to have a higher [Mn/Fe] ratio than the thick disk. 

The same authors also predict Ni to behave similarly to Mn. However, in Figure~\ref{fig:Fe-peak} the [Ni/Fe] remains  constant between thin and thick disk stars. More puzzling is the fact that [Ni/Fe] behaves the opposite at low metallicities, in which the Ni abundances of the \apoor accreted population are lower than the those of the \arich halo at same metallicities. This agrees with the results of \citet{Nissen2010}, but there is currently no explanation for this unexpected tendency. 

We note in {Figure \ref{fig:Fe-peak}} the very tight dispersion of stars within each population (i.e. stars which have the same symbols). This confirms that despite APOGEE data requiring some further calibrations of the absolute values of the elemental abundances, they provide good precision for Mn and Ni, demonstrating that this dataset is excellent for chemical tagging purposes. Therefore they are very good candidates to allow accurate distinctions between our populations, besides the fact that the comparison of the absolute abundances of [Mn/Fe] and [Ni/Fe] with external studies might be non trivial. 

\subsection{The Odd-Z, Even-Z, and Light Elements}
\label{subsec:oddevenlight}
\subsubsection*{C+N}
\begin{figure*}
\includegraphics[width=2\columnwidth]{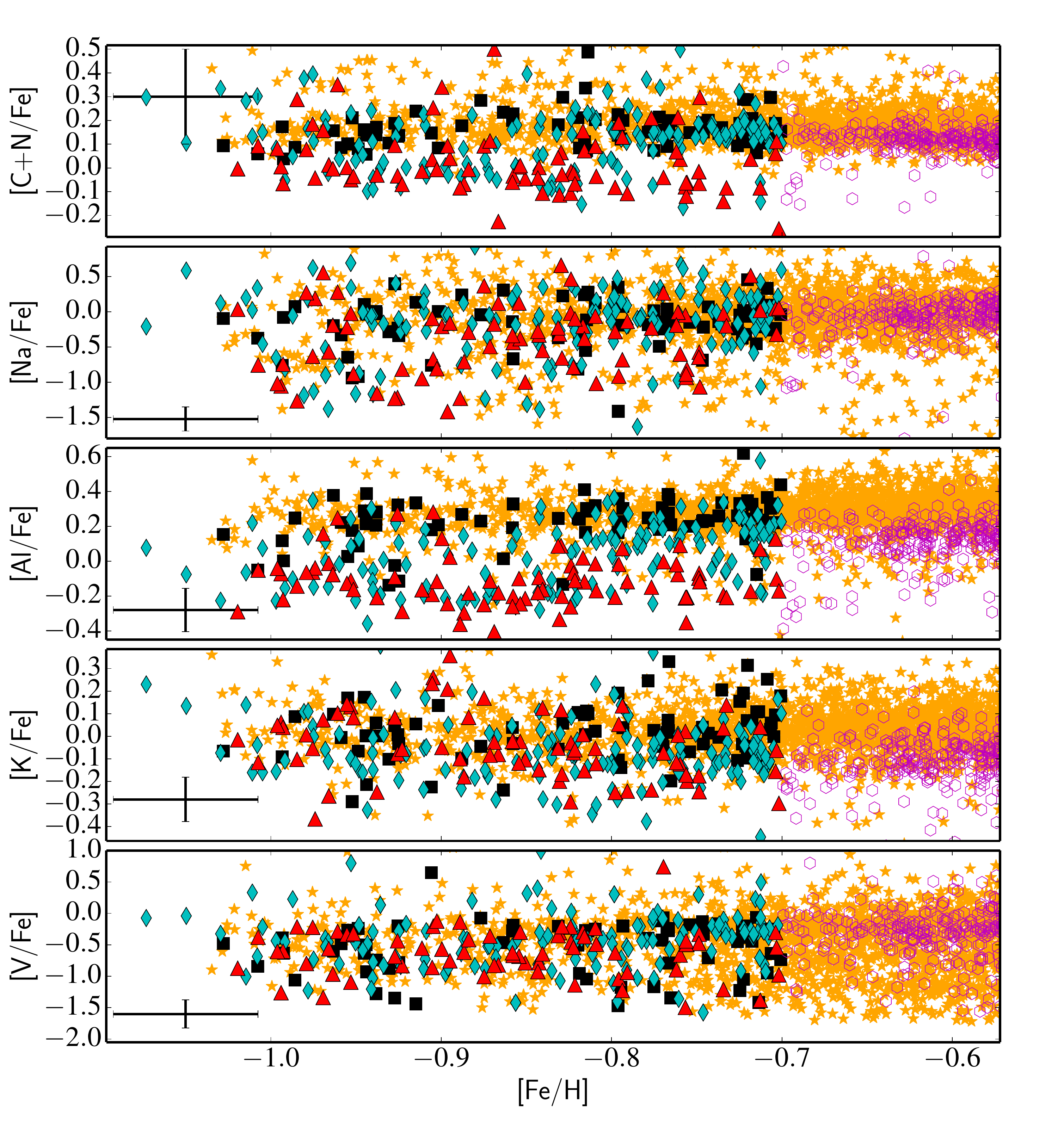}
\caption{The [C+N/Fe], [Na/Fe], [Al/Fe], [K/Fe], and [V/Fe] as a function of metallicity for the five subgroups are shown with same symbols as Figure~\ref{fig:alphachem}.}
\label{fig:oddevenlight}
\end{figure*}

APOGEE stars are giants, and as such their surface C and N abundances have been partially affected during their evolution as a result of dredge up processes \citep[e.g.][]{Iben1965}. However, because a large majority of the stars of this survey are expected to be low-mass stars ($\rm \approx 1M_\odot$), where CN cycling processes occurred, the initial C+N ratios are conserved throughout the evolution of those stars \citep[see][]{Masseron2015}. Therefore, we can use C+N to discuss and compare the evolution of C in our defined populations. In Figure \ref{fig:oddevenlight} we display in top panel [C+N/Fe] as a function of metallicity for our defined populations with the same symbols as previously.  \\
Carbon is mostly made by He burning and its main contributors into the enrichment of the Galaxy are SNII at very low metallicity as well as AGB stars at around --1.50 dex in metallicity. Hence, C+N is expected to globally increase with metallicity, but decrease as soon as SNIa kicks in because it does not produce C. This is why in Figure \ref{fig:oddevenlight}, we can observe some enhancement of [C+N/Fe] for the lowest metallicity stars, but then there is an overall decrease,  in particular in the thin disk which is the population most enriched by SNIa. Because of this unique interplay among other elements between different progenitors, and thanks to the good precision of C+N in the APOGEE data, the [C+N/Fe] ratios offer an opportunity to disentangle Galactic populations.

\subsubsection*{Odd-Z elements: Na, Al}
 The second and third panels of Figure~\ref{fig:oddevenlight} show the trends of [Na/Fe] and [Al/Fe] with metallicity, with the symbols representing the populations described in Sect.~\ref{subsec:kinematics}.  While Na abundances are relatively scattered at low metallicities, Al tends to separate. It is expected the Na and Al are produced in SNII. However, according to \citet{Kobayashi2006}, the production of those elements are strongly dependent on the initial C and N in the gas cloud that forms the stars. Therefore, it is expected that Na and Al are correlated with C+N. Besides the SNII production site, Na and Al are also expected to be partially produced by AGB stars in the Galaxy as shown by \cite{Nomoto2013} at the metallicities we focus on in this paper. This explains why in Figure~\ref{fig:oddevenlight} [Al/Fe] is rather enhanced in the lowest metallicity part of the thick disk population, because [C+N/Fe] is also enhanced. Moreover since SNIa do not produce Al and Na as efficiently as Fe, the [Al/Fe] and [Na/Fe] tend to decrease towards higher metallicities.
\citet{Nissen2010} demonstrate the effective ability of Na to characterize the \apoor populations. Unfortunately at these metallicities APOGEE has difficulty measuring Na \citep[see Figure 14 of][]{Holtzman2015}. Typical [Na/Fe] abundance uncertainties in this metallicity range approach $\sim \pm$ 0.20 dex. In contrast, the precision of Al abundances in the APOGEE data is effectively very high, and thus offers an alternative to Na.

\subsubsection*{K and V}
The last bottom panels of Figure~\ref{fig:oddevenlight} display the trends of potassium and vanadium as a function of metallicity for our different defined populations with different symbols.   The abundance V shows a very large dispersion and remains rather constant among all populations, thus tracking Fe. This is similar to what is seen in the literature for V \citep[e.g.][]{Bodaghee2003, Battinstini2015}. Concerning K, the abundances of this element seem to decrease towards higher metallicity consistent with literature \cite[e.g.][]{Shimansky2003}. One can also notice a decrease in some of the accreted population stars, but this trend is rather shallow. Although we note that the nucleosynthesis channel is rather misunderstood for K and V, and the supernovae yields lead to largely underestimated values compared to their observed Galactic abundances \citep[e.g.][]{Nomoto2013}. As such, these elements do not give strong constrain to distinguish our defined populations.
\\
\\
With the element-by-element discussion above, in the next section we focus on using the abundance ratio trends to inform our understanding of the transition between the Galactic disks and halo. 

  \section{Implications of chemical abundance trends}
In the previous section, we have demonstrated that among $\alpha$-elements, O and Mg bear distinction of the star formation activity, particularly at low metallicities. It also indicates that Mn and Ni are potential alternatives to disentangle the signatures of SNIa. Regarding the accreted population versus thin disk, Al and C+N represents a good diagnostic as they contain signatures of AGB nucleosynthesis. The observations of those elemental abundances provide insight about their origin, but this requires some further modeling taking into account stellar yields as well as a thoughtful discussion of the observational bias that is beyond the scope of this paper. Nevertheless, we will rather use their relative abundance ratios in this section to discuss the connection and the transition phases of the different canonical components of the Milky Way. 
\label{sec:discussion}

\subsection{Thick disk-halo transition}
\label{subsec:transition}
\begin{figure*}     
 	 \includegraphics[width=2\columnwidth]{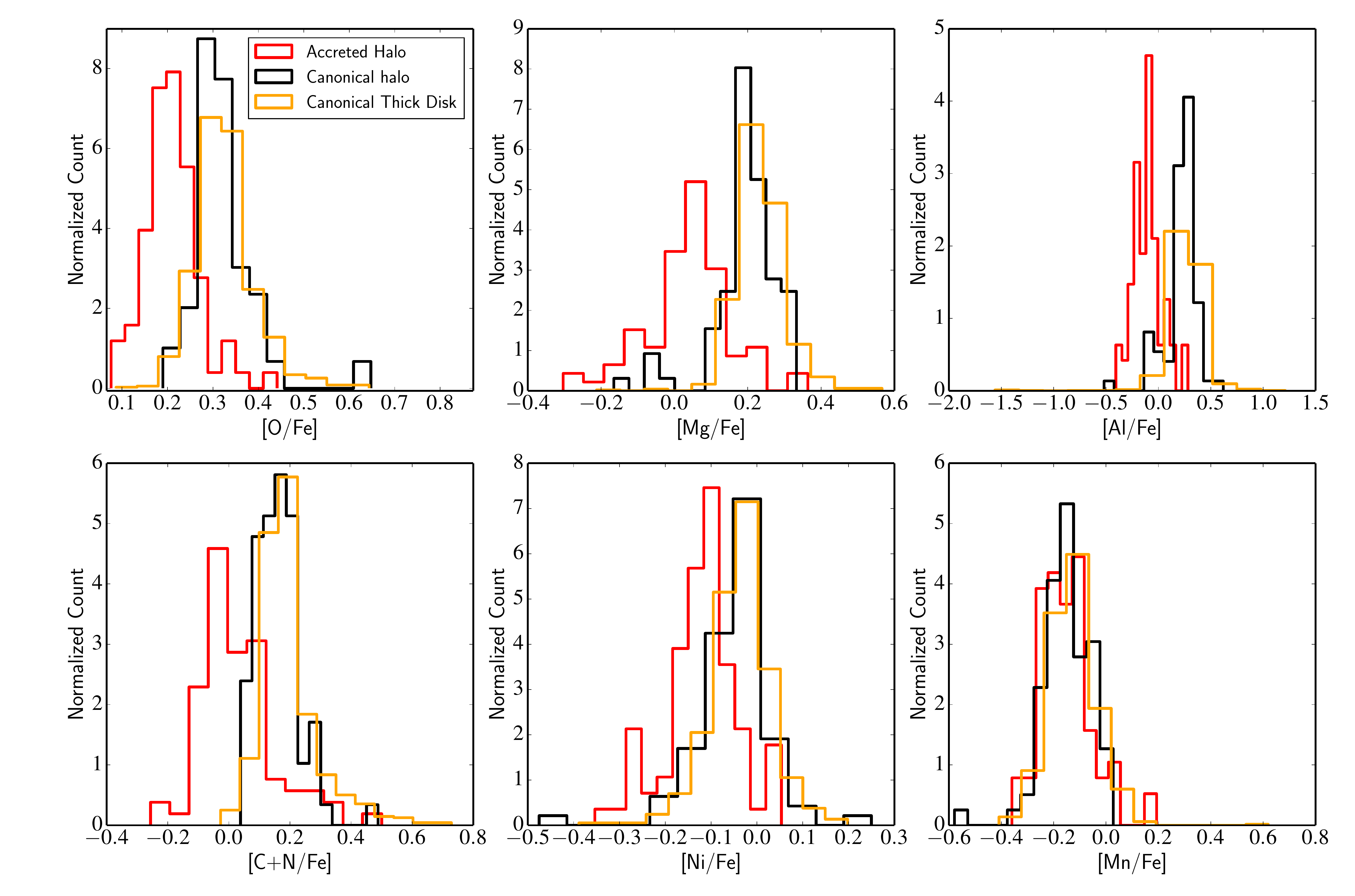}

	\caption{The distribution of [O/Fe],  [Mg/Fe],  [Al/Fe], [C+N/Fe], [Ni/Fe], [Mn/Fe] abundance ratios for the accreted halo (red), canonical halo (black) and canonical thick disk sequence (orange).  We find that in most elements (except Mn, Na) are systematically under abundant in the accreted halo sequence relative to the canonical halo sequence. The abundance dispersion in the accreted halo sequence is comparable or in most cases larger than the canonical halo sequence. Further Mn shows no chemical difference between the two sequences. In all chemical abundance ratios shown (and those not shown as well) the canonical thick disk and halo components overlap significantly. }
	\label{fig:chemdist_apoor}
  \end{figure*}

One of the primary goals of this work is to use the chemistry of a large sample of stars to study the disk-halo transition in the metallicity region where they are thought to overlap (--1.20 $<$ [Fe/H] $<$ --0.70 dex) based on literature (see section \ref{subsec:decomp_intro} for more details). In each element studied, the canonical halo (black squares) and canonical thick disk (orange stars) components are completely indistinguishable (see Figures \ref{fig:alphachem}, \ref{fig:Fe-peak}, \ref{fig:oddevenlight}). This result was tested using different GRV dispersion cuts between 1 -- 4$\sigma$ (i.e. varying the amount of contamination between the two groups) and in all cases the canonical thick disk and halo overlap in every element. Another illustration of this can be seen in Figure \ref{fig:chemdist_apoor} for the key elements we have previously highlighted. In each panel of that figure the canonical thick disk (shown in orange) and the canonical halo (shown in black) \textit{not only overlap but have comparable dispersions} suggesting that the two formed in a similar high star formation environment with well-mixed turbulent gas making them indistinguishable in chemistry leaving only the kinematics and spatial distributions to separate the two. 

\cite{Ishigaki2012} found, using a sample of 97 stars, evidence that [Mg/Fe], [Si/Fe] in the thick disk are higher, on average, than halo stars. Their follow-up study \citep{Ishigaki2013} found that [Na/Fe] and [Ni/Fe] were lower in the halo compared to the thick disk. It is important to note that we select our halo samples in different ways making it difficult to compare these two studies. The halo sample in \cite{Ishigaki2012} was selected kinematically on the basis of the Toomre diagram and is split into an `inner halo' and `outer halo' component. It is likely that both of their inner and outer halo components contain stars that we would classify as \arich and \apoornospace. In our study, we selected the halo sample kinematics (using the $l$-GRV diagram) and found a bimodal distribution in \afenospace. As a result of this we called the \arich components the `canonical halo' and the \apoor component the `accreted halo'  \citep[based on studies such as][]{Nissen2010}. Since both halo samples in \cite{Ishigaki2012} likely contain \arich and \apoor stars, the mean \afe could be driven to lower values which may explain why the [Mg/Fe] and [Si/Fe] may have been found to be on average lower in the halo(s) than the thick disk.

The only chemical distinction between thick disk and canonical halo is solely in the mean and dispersion of the metallicity distribution function for the two components. This result could be biased by the fact that we choose the canonical halo to be \arichnospace, similar to the thick disk. It is interesting to point out that while we selected the canonical halo and thick disk to be \arichnospace, they are, in fact, similar in all other elements including elements that have different production channels to \afe (e.g. odd, even, and light elements) that we do not use for selection. If we had not initially separated the halo sample into \arich and \apoor subsamples, we would conclude that in key elemental abundance spaces the halo is bimodal (see Figure \ref{fig:chemdist_apoor}) and the \arich component of that bimodal distribution is chemically similar to the thick disk despite having a different metallicity distribution function while the \apoor component is chemically different. It is also interesting to note that the canonical halo has some evidence of spatial-kinematic coherence (black squares in Figure \ref{fig:lGRV}), indicating it may have net rotation in line with other studies \citep[e.g.][]{Carollo2007, Carollo2010}. This may suggest that the thick disk-halo transition is smooth kinematically and would be worth examining with full 3D space motions to confirm. As the thick disk is thought to form from well-mixed turbulent gas \citep[e.g.][]{Haywood2013}, we interpret this smooth chemical transition between the canonical thick disk and halo components as evidence that the gas of the inner in situ Galactic halo may have been the precursor to the thick disk. We note that subject to careful modeling of selection function, most of the inner halo (i.e. the \arich stars) chemically resembles a single homogeneous self-enriching population, in which the major transition between the halo and thick disk is pressure support to angular momentum support in kinematic space.

\subsection{The accreted halo}
\label{subsec:apoorhalo}
In section \ref{subsec:kinematics} we showed that there is a sequence of \apoor metal-poor halo giant stars in APOGEE (which we designated as `accreted halo')  that resembles the \apoor sequence of \cite{Nissen2010}, \cite{Ramirez2012}, \cite{Bensby2014}, \cite{Hawkins2014}, and others. What makes this result interesting is that APOGEE has a very different selection function and surveys a larger volume than past studies \citep[e.g.][]{Nissen2010} which primarily rely on chemical analysis of relatively local dwarf stars. The discovery of the \apoor sequence in a large sample of giant stars from APOGEE with a different selection function and larger survey volume strengthens the argument that the \apoor halo sequence is real.

We used the APOGEE survey to study the chemistry of these \apoor stars (red triangles) in section~\ref{sec:chemistry} by comparing their abundances in 14 elements, including $\alpha$ (Figure~\ref{fig:alphachem}), Fe-peak elements (Figure~\ref{fig:Fe-peak}), odd, even, and light elements (Figure~\ref{fig:oddevenlight}) to the canonical halo sequence (black squares).  We found that the accreted halo is primarily distinguishable in the $\alpha$-elements (e.g. Mg, O, and S). To further illustrate the chemical distinction between the accreted halo and canonical halo sequences, in Figure~\ref{fig:chemdist_apoor} we plotted the chemical distribution of six key elements from all three elemental groups for the canonical halo (black line),  accreted halo (red line) components. 

In all chemical elemental abundance ratios shown in Figure \ref{fig:chemdist_apoor} except [Mn/Fe] the \textit{accreted halo is systematically under abundant compared to the canonical halo.} The lower [C+N/Fe] of the accreted halo indicates that the initial chemical composition of the gas that formed the \apoor stars were different from the \arich providing further evidence that the \apoor stars are likely extragalactic in origin \citep[e.g.][]{Nissen2014}. This is the first observation that [Al/Fe] is under abundant in the accreted halo population. This result is not surprising as Al is thought to be produced in SNII and correlated strongly to C+N. Therefore, if [C+N/Fe] is under abundant \citep[see section \ref{subsec:oddevenlight} and][]{Nissen2014} one would expect [Al/Fe] to also be under abundant. Furthermore, the dispersion of the accreted halo in all elemental ratios except Mn is larger than the canonical halo component. This may suggest several extragalactic systems being responsible for the production of the accreted halo component in contrast to the canonical halo, which is thought to form from one homogenous gas cloud. However, we remind the reader that the dispersion in each element is sensitive to the boundary that separates the canonical and accreted halo components. As such, we tested various boundaries between the canonical and accreted halo components to study how this would affect the dispersions and the separation of the two components.  We shifted the line that separates the \arich canonical and \apoor `accreted' halos by at least $\pm$0.05 dex (twice the typical internal uncertainty in \afenospace). Lowering the boundary causes stars currently classified as accreted with moderate \afe\ to be reclassified as canonical halo. The overall result is a decrease in the dispersion of accreted halo in most elemental spaces and an increase in the dispersion in the elemental distributions for the canonical halo. Raising the boundary has the opposite effect. In both cases, the accreted halo is chemically distinct from the canonical halo in the elements listed above. If we were to not separate the halo into a `canonical' and `accreted' component at the onset, we would find a bimodal distribution in O, Mg, Al, C+N and Ni in the full halo sample consistent with two chemically distinct components.

As noted by \cite{Nissen2011} the [Mn/Fe] ratio is comparable between the \apoor and \arich halo sequences. This is likely a result of the metallicity dependence on the SNIa yields. We also found that [K/Fe] and [V/Fe] ratios are comparable both in mean and dispersion between the accreted and canonical halos indicating that these elements do not give strong constraints to distinguish the canonical and accreted halo components. We recall that the [V/Fe] ratios have significantly larger scatter than expected from other literature \citep[e.g.][]{Fulbright2000, Battinstini2015} which may indicate this element is less reliable. 

Additionally, increasing the kinematic GRV dispersion cut to 4$\sigma$ reduces the contamination of the disk population in the halo stars however decreases the sample size significantly while reducing the kinematic GRV dispersion cut to 1$\sigma$ increases the contamination from the \apoor disk population thereby broadening the chemical abundance ratio dispersions of the \apoor `halo' sequence. In all cases in which we tested various GRV dispersion cuts, the canonical and accreted halo remain chemically distinct in the [C+N/Fe], [Al/Fe], [Ni/Fe] and the $\alpha$-elements (particularly [O/Fe], [Mg/Fe], [S/Fe]). 

Simulations of the Galactic halo \citep[e.g.][]{Bullock2005, Zolotov2009, Cooper2010} have indicated that the halo may have been constructed primarily through a combination of dissipative collapse and accretion events of massive systems. The likely accreted \apoor stars represent accretion from systems onto the halo and their \afe and chemical abundance distributions may be able to give insight to the mass of the systems that accreted. Furthermore, a careful study of the relative fraction of the accreted \apoor halo stars and the in situ \arich halo stars will provide unique constraints on the relative importance of accretion events in the assembly of the Galactic halo. With the upcoming surveys it will be possible to build large samples of \apoor halo stars that will make these tests possible.  

\subsection{The undetermined group: thin disk at low metallicities?}
\label{subsec:thindisk}
The thin disk is thought to be rather metal-rich with [Fe/H]~$>-0.70$~dex \citep[e.g.][]{Reddy2003, Fuhrmann2004}. However, there has been some debate as to whether the thin disk actually extends to metallicities as low as [Fe/H] =~--1.0 dex \citep[e.g.][]{Mishenina2004}. Most recently \cite{Bensby2014} used a sample of 714 F and G dwarf stars and found that the lower metallicity for the thin disk was [Fe/H] $\sim$~--0.70~dex. Their finding in combination with the extent of the accreted halo \citep[e.g.][]{Nissen2010} is the reason our initial cut in \feh\ for our sample was at \feh~=~--0.70~dex separating the accreted halo from the canonical thin disk in the \afenospace-metallicity diagram. Upon separating our metal-poor sample into several subpopulations in section \ref{subsec:kinematics} we found a non negligible amount of stars that are metal-poor, \apoor yet the kinematics suggest they may be disk-like, which we called the undetermined subgroup (cyan diamonds).

\begin{figure*}
 \subfigure[]{
              \includegraphics[width=2\columnwidth]{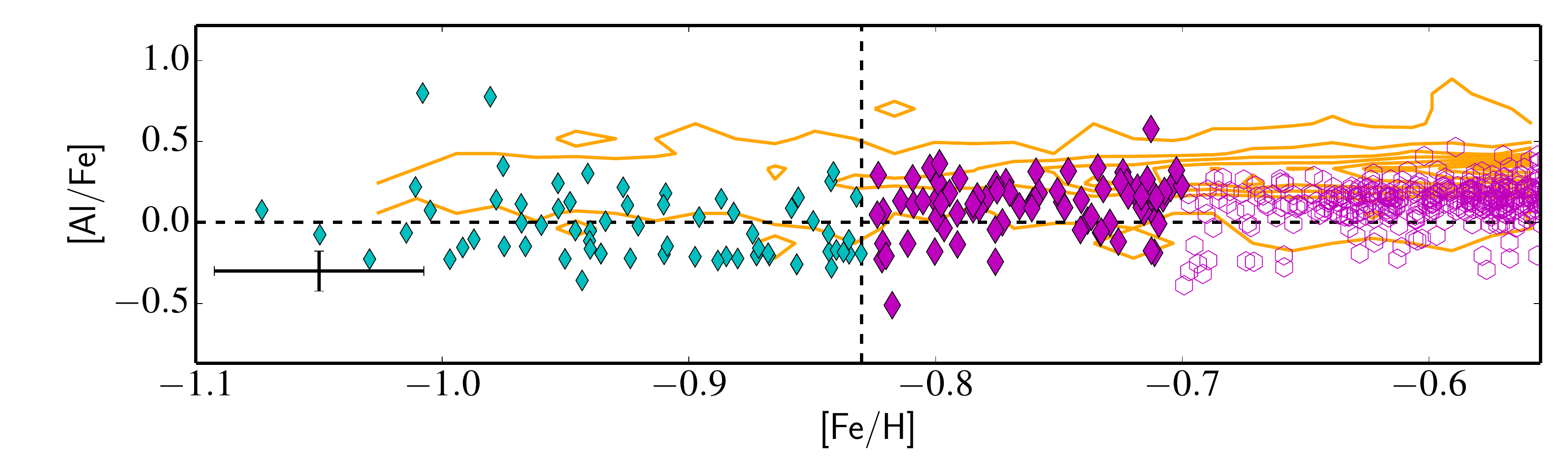}
                \label{fig:ALFE_mpthin}  }
        \subfigure[]{
              \includegraphics[width=2\columnwidth]{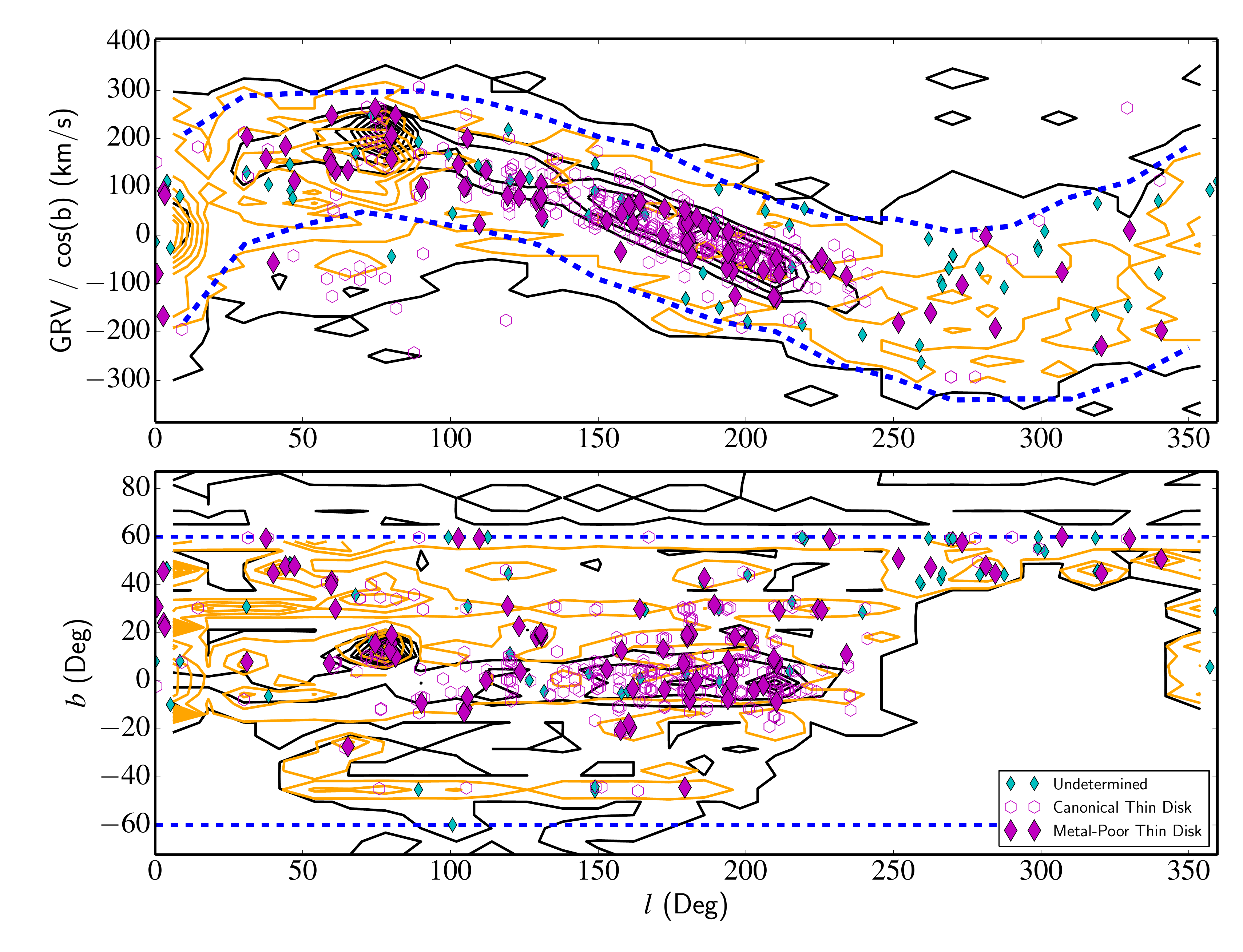}
               \label{fig:lGRV_mpthin}   }

 \caption{(a) [Al/Fe] as a function of metallicity of the candidate metal-poor thin disk (magenta diamonds). All \arich stars are shown as orange contours. The canonical thin disk sample is shown as open magenta hexagons for comparison. Dotted lines are placed at the metallicity cut \feh\ = --0.83 dex and a guiding line is placed at [Al/Fe] = 0 dex. (b) Upper panel: The GRV as a function of $l$ for the candidate metal-poor thin disk. The full APOGEE sample that pass our quality cuts are shown as black contours. For consistency the blue dotted lines represent the initial kinematic selection. Lower Panel: The Galactic coordinates ($l$, $b$) of the various subgroups following the same symbols as the top panel. We found that the candidate metal-poor thin disk with metallicities down to $\sim$ --0.83 dex have kinematic-spatial coherence that is consistent with the canonical thin disk at these metallicities. } 
  \label{fig:chem_lGRV_mpthin}
  \end{figure*}

In Figure \ref{fig:oddevenlight}, it is clear that the [C+N/Fe] and [Al/Fe] abundances of the undetermined subgroup at the high metallicity end (~--0.83 $<$ [Fe/H] $<$ --0.70~dex) are, on average, consistent with the thin disk population at those metallicities while at lower metallicities ([Fe/H] $<$ -- 0.90 dex) the undetermined population decreases in both [C+N/Fe] and [Al/Fe] and behaves like the accreted halo subgroup. For this reason we changed the colour of the diamonds with metallicity \feh\ $>$ --0.83 dex to magenta in order to denote that these stars chemically resemble the canonical thin disk population. In Figure \ref{fig:ALFE_mpthin}, we plot [Al/Fe] as a function of metallicity now with all \arich stars as contours with the canonical thin disk (open magenta hexagons) and our candidate metal-poor thin disk (magenta diamonds). We leave the diamonds with metallicity \feh\ $<$ --0.83 dex as cyan because these stars are still undetermined although it is likely they are either accreted halo stars or miscategorized thick disk stars.   

In the top panel Figure \ref{fig:lGRV_mpthin} we plot the GRV/cos($b$) as a function of $l$ and $b$ as a function of $l$ in the bottom panel for emphasizing the candidate metal-poor thin disk (magenta diamonds) and undetermined (cyan diamonds) subgroups relative to the canonical thin disk component (open magenta hexagons). We plot all \arich stars (i.e. canonical thick disk and halo) as background orange contours, following our notation color for the thick disk, for comparison. We note the following: (1) the canonical thin disk at intermediate metallicities (open magenta hexagons) population is predominately found between $90 < l < 220$ deg and (2) the candidate metal-poor thin disk (magenta diamonds) are spatially and kinematically consistent with the canonical thin disk population. Specifically,  the GRV dispersion of the candidate metal-poor thin disk at a constant $l$  (i.e. magenta diamonds in Figure \ref{fig:lGRV_mpthin}) is very small and is thus consistent with thin disk. This suggests that the \textit{canonical thin disk may extend to lower metallicities than found in recent studies} \citep[e.g.][]{Bensby2014}. The lower metallicity extent compared to other studies \citep[such as][]{Bensby2014} could be a result of the fact that APOGEE has a large sampling of stars in the outer Galaxy, where the most metal-poor thin disk stars should exist. As noted in section \ref{subsec:data}, while the precision of the metallicities are high, the accuracy or metallicity scale may not be well calibrated. \cite{Holtzman2015} used a sample of globular clusters and found that in this metallicity regime (\feh\ = --0.6 to --1.0 dex),  [Fe/H] can be overestimated by as much as $\sim$ 0.10 dex. This causes the metallicity extent of the thin disk to extend down to at least $\lesssim -0.83$ dex if we use the uncalibrated [Fe/H] or --0.95 dex if one calibrates the [Fe/H] scale using the globular clusters. We remind the reader we choose to use the uncalibrated [Fe/H] because the [X/Fe] trends match what is expected from literature while calibrating the [Fe/H] values has the effect of inflating the [X/Fe] \citep{Holtzman2015}. Given this uncertainty in the calibration, we plan to explore the metal-poor extent of the thin disk with other surveys, such as the Gaia-ESO survey \citep{Gilmore2012}, in order to investigate the absolute value of the metal-poor tail of the thin disk metallicity distribution. 

Another interesting result to point out from Figure \ref{fig:lGRV_mpthin} is that our candidate metal-poor thin disk (magenta diamonds) and most of the intermediate metallicity stars of the canonical thin disk, are found at locations outside the solar circle ($90 < l < 270$ deg) while the more metal-rich canonical thin disk is seen at Galactic longitudes consistent with locations inside the solar circle indicative of a negative radial-metallicity gradient. Negative metallicity gradient are widely observed in the canonical thin disk both in small samples \citep[e.g.][]{Genovali2014} and large surveys \citep[e.g.][]{Cheng2012, Mikolaitis2014}. Theoretically, the presence of a negative metallicity gradient in the canonical thin disk is evidence for an ``inside out" formation scenario of the disk. Under this scenario, the oldest disk population is formed from the low angular momentum gas, which falls into the center of the halo and begins rapidly forming stars earlier compared to the outer disk. The rapid star formation leads to larger metallicities being produced in the central disk compared to the outer disk causing a negative metallicity gradient \citep[e.g.][]{Larson1976, Minchev2014b}. Our finding is thus suggestive of a negative radial metallicity gradient, which is consistent with the inside-out formation scenario.

\section{Redefining the selection of Galactic components: a chemical tagging approach}
\label{sec:chemtag}
The current method of decomposing the canonical Galactic component in the current scheme of chemokinematics (the use of both chemistry, e.g. \afe and metallicity) and kinematics (e.g. radial velocities, or U,V,W velocity vectors) has the disadvantage that it can be quite complex. The complexities lie in controlling the assignment of stars in the populations. For example, we saw in section \ref{subsec:kinematics} where we had to use both chemistry and kinematics just to separate out the Galactic components and even had an extra undetermined population that could not be categorized until after full chemical analysis. This caused each plot to contain many more symbols (and thus populations) than may be necessary. Furthermore, the assignment of populations on just kinematics \citep[e.g.][]{Bensby2014} or both chemistry and kinematics causes significant biases making it difficult to discern the chemical evolution of the Galaxy. This leads us to question: Is there a more simple approach? More specifically, can we just use chemistry to directly separate out the canonical Galactic components to then study the spatial and dynamical distributions?

 \begin{figure*}
  \includegraphics[width=2\columnwidth]{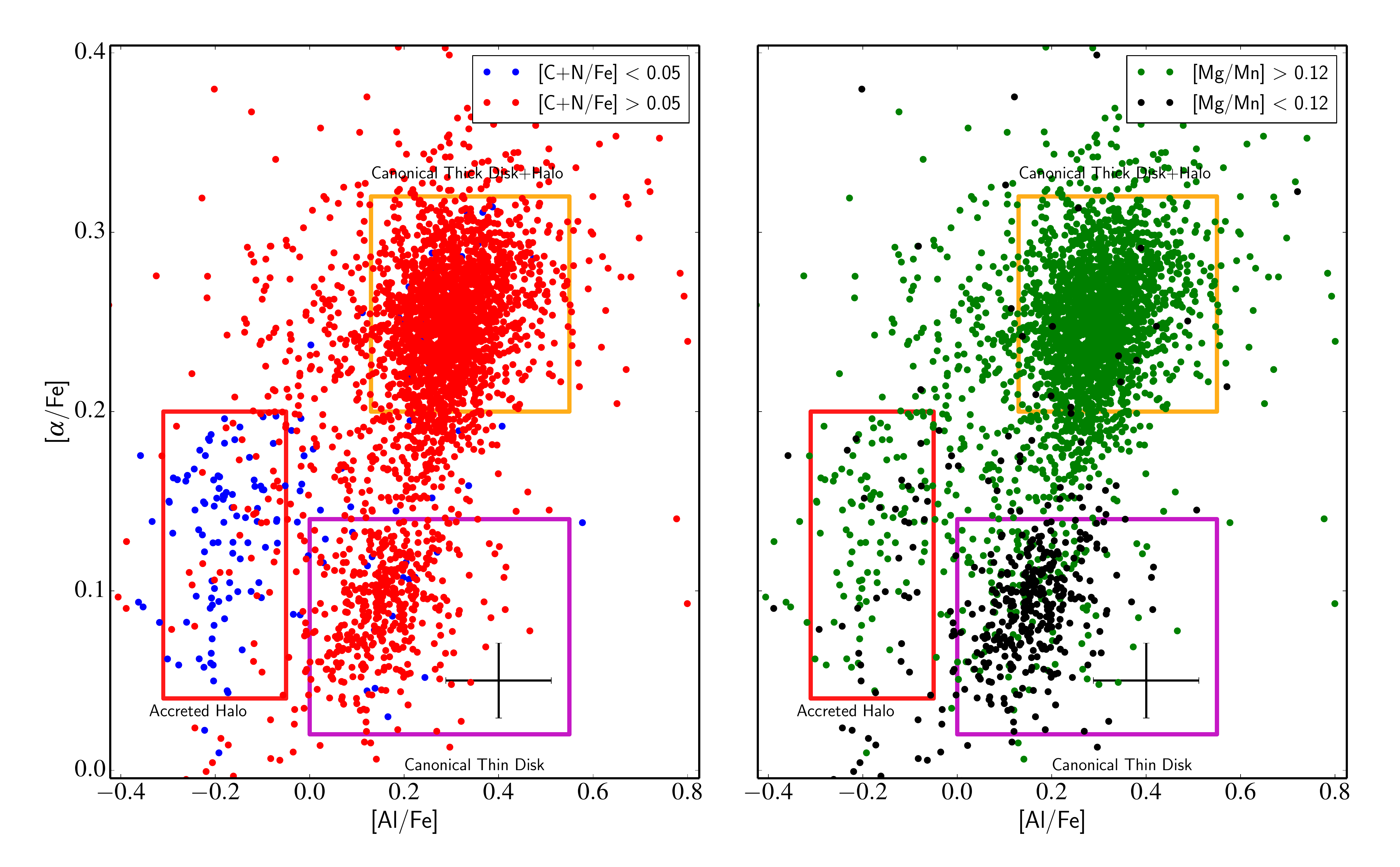}
 \caption{Left Panel: The \afe abundance ratio as a function of [Al/Fe] for all intermediate metallicity stars with --1.20 $<$ \feh\ $<$ --0.55 dex coloured by [C+N/Fe] where red is [C+N/Fe] $>$ +0.05 dex and blue is [C+N/Fe] $<$ +0.05 dex. Right Panel:  The same as the right panel but now colour coded by [Mg/Mn] where green is [Mg/Mn] $>$ +0.12 dex and black is [Mg/Mn] $<$ +0.12 dex. In both panels, the error bar represents the typical uncertainty in the abundance ratios. There are three clusters within the data. The canonical thick disk + canonical halo, canonical thin disk and accreted halo components which should fall within the orange, magenta and red selection boxes, respectively. Note with just these four chemical dimension each component is marked by a different colour (e.g. accreted halo it is blue indicating $\alpha$-poor, Al-poor, and [Mg/Mn] is mixed while the canonical thin disk is  $\alpha$-poor, Al-rich and [Mg/Mn] is low) and thus will very strongly separate in these chemical spaces.} 
  \label{fig:chemtag}
  \end{figure*}

In this section we attempt to answer this question by putting forward a chemical-only labelling approach to separate the Galactic components based on the above analysis. In section \ref{sec:chemistry} and \ref{sec:discussion}, we discussed that among the various elements studied here the canonical halo, accreted halo and thin disk were able to be separated primarily in the $\alpha$-elements (e.g. \afe, Mg, or O), Al, C+N. In this way, we can use a combination of these elements relative to Fe (or even other Fe-peak elements such as Mn) to chemically separate the canonical thin disk, accreted halo and thick disk. We note from above that the canonical thick disk and halo are effectively chemically indistinguishable meaning they can, in principle, be treated as one population although they likely have an inhomogeneous kinematic and spatial distribution.

The \afe ratio is particularly important because it tracks the ratio of the elements produced in SNII and SNIa providing useful constraints on the formation timescale of the component. We note here that the abundance ratio [Mg, O/Mn] would actually be a better tracer of SNII/SNIa as Mg and O are the first elements to be affected as a result of SNII \citep[e.g.][]{Nomoto2013} and Mn is produced at higher fractions compared to Fe in SNIa \citep[e.g.][]{Gratton1989}. However, we choose to use \afe in part because it is more accessible in other non-APOGEE data compared to Mn as well as the fact that Mn is difficult to measure in optical spectra as several lines suffer from hyperfine structure splitting \citep[e.g.][Jofr\'e et al., submitted]{Nissen2000}. Al is important as it is produced via SNII and is sensitive to the initial C and N abundance, which in turn is produced by He burning or AGB evolution. For these reasons, in Figure~\ref{fig:chemtag} we show \afe as a function of [Al/Fe] in both panels. The data colored by their [C+N/Fe] in the left panel (red dots have [C+N/Fe]~$>$~+0.05  dex while blue dots have [C+N/Fe]~$<$~+0.05~dex) and [Mg/Mn] in the right panel (green dots have [Mg/Mn]~$>$~+0.12  dex while black dots have [Mg/Mn]~$<$~+0.12~dex). The data separates in three clusters in Figure~\ref{fig:chemtag}. One being the thin disk stars with low \afenospace, moderate [Al/Fe] and low [Mg/Mn] shown by a magenta selection box. The second is the thick disk+halo stars that have higher \afenospace, higher [Al/Fe] and high [Mg/Mn] with an orange selection box. Lastly is the accreted halo stars that have low \afe with a larger spread than the thin disk, low [Al/Fe] and moderate to low [Mg/Mn] with a red selection box.

\begin{figure*}     
       \subfigure[]{
              \includegraphics[width=2\columnwidth]{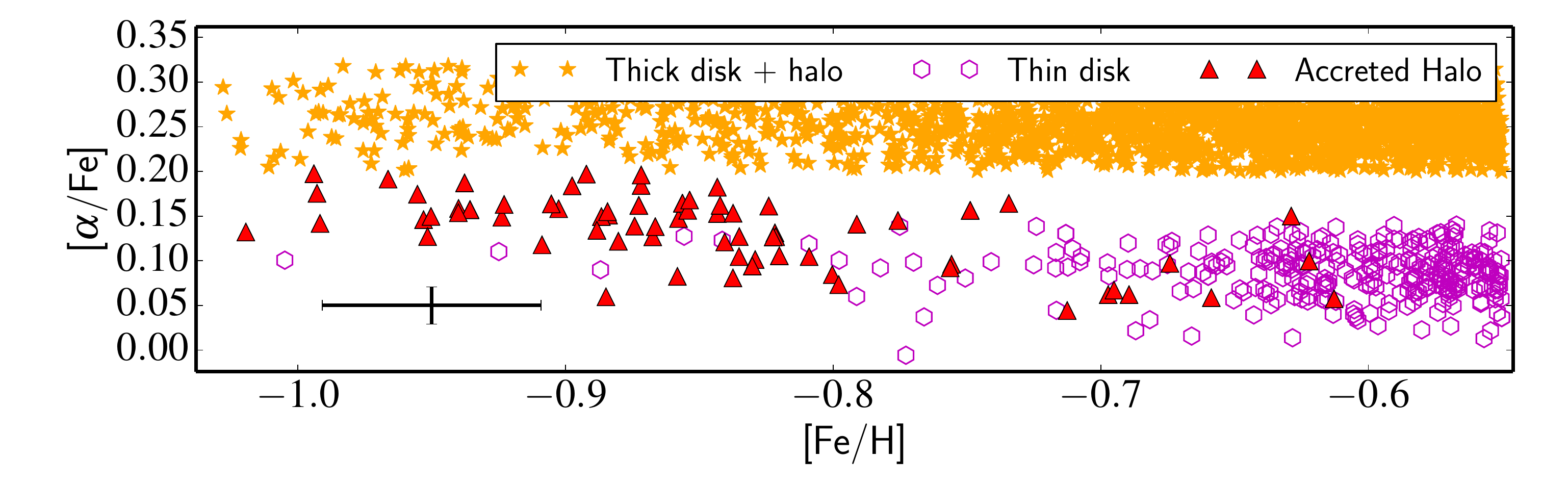}
                \label{fig:alphamet_chemtag}  }
        \subfigure[]{
              \includegraphics[width=2\columnwidth]{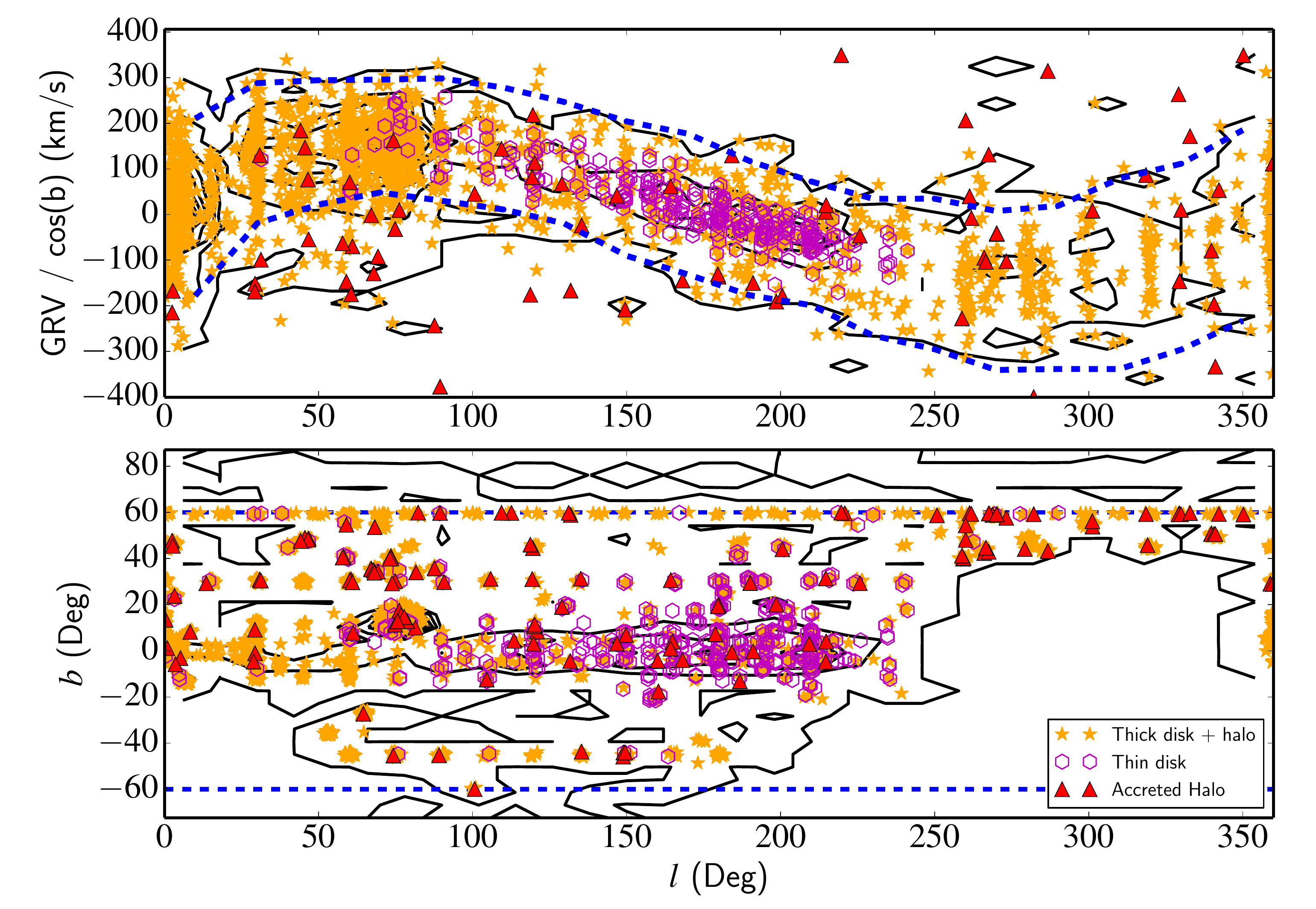}
               \label{fig:lGRV_chemtag}   }
	 \caption{(a) The \afe as a function of metallicity for the intermediate metallicity sample using only chemistry to separate the components. The canonical halo + thick disk stars (orange selection box of Figure \ref{fig:chemtag}) are shown as orange stars.  The canonical thin disk (black dots in the magenta selection box of Figure \ref{fig:chemtag}) are shown as open magenta hexagons. The accreted halo (blue dots in the red selection box of Figure \ref{fig:chemtag}) are shown as red triangles. (b) Upper panel: The GRV as a function of $l$ for the intermediate metallicity sample using only chemistry to separate the components with the same symbols as (a).  For consistency the blue dotted lines represent the initial kinematic selection. Lower Panel: The Galactic coordinates ($l$, $b$) of the various subgroups. } 
  \label{fig:alphamet_lGRV_chemtag}
  \end{figure*}

As a consistency check of the populations defined from the selection boxes in Figure~\ref{fig:chemtag}, we looked at where they fall in the \afenospace-[Fe/H] plot (Figure \ref{fig:alphamet_chemtag}). In Figure \ref{fig:alphamet_chemtag} we see that we recover the same structure seen in Figure \ref{fig:alphamet}. The canonical halo and thick disk are not chemically distinct and represent the \arich stars. The thin disk is \apoor and primarily at higher metallicities but extend to very low metallicities and the \apoor accreted halo is primarily found at low metallicities although extend to as high as --0.60 dex consistent with \cite{Nissen2010}. The large gaps in the \afe in Figure~\ref{fig:alphamet_chemtag} are artificial as they are a result of our selection boxes. We plan to explore a more sophisticated statistical approache to improve or abandon the selection boxes in favor of a probabilistic and clustering approach. Here we show that in a simple case of using \afenospace, [Al/Fe], [C+N/Fe] and [Mg/Mn] we can indeed chemically tag the bulk of the Galactic components in a robust way.

This is confirmed when looking at the $l$-GRV space (Figure \ref{fig:lGRV_chemtag}). We use $l$-GRV space to determine if this purely chemical definition of the canonical Galactic components recover the kinematic structure we expect them to have. Like in Figure \ref{fig:lGRV}, the dotted blue line represents the sinusoidal pattern of stars following disk-like motion. The canonical thin disk (open magenta hexagons) not only have spatial-kinematic coherence indicating co-rotation with the disk, they are also located primarily outside of the solar circle. The accreted halo (red triangles) have hot kinematics although some are located inside a region of $l$-GRV space where they may be co-rotating with the disk (i.e. inside the blue dotted line in Figure \ref{fig:lGRV_chemtag}). Full 3D kinematics would be very helpful to validate these stars as accreted halo. In addition, the accreted halo may extend to metallicities as high as \feh\ $< -0.6$ dex. The canonical thick disk + halo (orange stars) are dominated at these metallicities by the thick disk and thus by in large show a co-rotation with the disk. The GRV/cos($b$) dispersion is much larger than the thin disk as expected. However there are orange stars that reside quite far outside the disk-like patter (blue dotted line) which are the canonical halo stars. This test verifies that by \textit{using the \afenospace-[Al/Fe] diagram in connect with [C+N/Fe] and [Mg/Mn] we can distinguish Galactic populations in a way that can be studied free of biases in the dynamical spaces.} This has the advantage to traditional methods that full 3D kinematics are not necessary (outside of validation) and one can break the degeneracies in the \afenospace-metallicity diagram (i.e. the initial undetermined group is now categorized, see Figure \ref{fig:alphamet_chemtag}). We also have less contamination in that we no longer have thin disk stars that have extreme velocities (i.e. the open magenta hexagons outside the dotted blue line in Figure \ref{fig:lGRV}). The population assignment also works just using the \afenospace-[Al/Fe] diagram alone rather than adding the addition [C+N/Fe] and [Mg/Mn] abundance ratios. The advantage to adding those ratios is to reduce contamination from the fringes of the thick disk and thin disk components. This is just a first step in showing that chemical-only approaches may be the way forward in decomposing the Galaxy especially with large chemical surveys either underway or planned for the near future. More work will be needed to move beyond the selection boxes we use here to more robustly quantify the discreetness between the components in these chemical spaces, which we plan to explore in a forthcoming paper.

\section{Summary}
\label{sec:summary}
We have studied the chemistry of a relatively large sample of intermediate metallicity ([Fe/H] $<$~--0.55~dex) giant stars using the APOGEE survey in order to explore the disk-halo transition though a chemical tagging approach. We selected a sample of $\sim$3200 intermediate metallicity stars using a set of quality control cuts described in section \ref{subsec:data}. We used the \afenospace-\feh\ diagram to split and $l$-GRV diagram \citep[e.g.][]{Majewski2012, Sheffield2012} to split our sample into five subgroups: (1) canonical thin disk (open magenta hexagons), (2) canonical thick disk (orange stars), (3) canonical halo (black squares), (4) accreted halo (red triangles), (5) undetermined (cyan diamonds) as described in section \ref{subsec:kinematics}. In section \ref{sec:chemistry} we discussed the chemical abundance patterns of these four groups in $\alpha$-elements (Mg, Ti, Si, Ca, O, S), Fe-peak elements (Mn, Ni), and odd, even, and light elements (C+N, Na, Al, K, V). 

Our main results of this Galactic chemical evolution analysis can be summarized in the following points: 
\begin{itemize}

\item The canonical halo and canonical thick disk are not chemically distinct in any of the elements studied. This is most evident in the abundance ratio distribution in the various elements shown in Figure \ref{fig:chemdist_apoor} which indicates that the \arich halo component (black line) and the \arich disk component (orange line) overlap both in terms of mean and dispersions for both populations. This may suggest that the Galactic halo and thick disk are chemically formed from similar gas but one is pressure supported and the other is angular momentum support. 

\item We uncovered the largest sample of `accreted' halo stars. We showed, for the first time, that accreted halo stars are under abundant in [S/Fe], which is unsurprising given its status as an $\alpha$-element. We have also showed that the accreted halo is under abundant in [Al/Fe] which is likely caused by the nuclear pathways in which Al is produced. The low abundances of [C+N/Fe] of accreted halo stars respect to canonical stars imply that this gas had a slower chemical environment \citep{Nissen2014}. The Ni of the \apoor sequence is, on average, under abundant relative to the \arich sequence consistent with previous studies. These results indicate that we have confirmed the existence of an \apoor halo sequence of stars that is chemically different than the stars formed the rest of the halo with chemistry suggesting that these stars may born externally \citep[e.g.][]{Nissen2010, Nissen2011, Schuster2012, Ramirez2012, Hawkins2014}. 
 
\item Upon closer investigation of the undetermined population, we have found evidence that the \textit{thin disk extends down to very low metallicities ([Fe/H] $\lesssim~-0.83$ dex)}. In Figure \ref{fig:lGRV_mpthin}, we showed that the candidate metal-poor thin disk, which was the metal-rich end of the undetermined population, were consistent with the canonical thin disk at comparable metallicities both kinematically and spatially. This was the first indication that these stars may be among the lowest metallicity thin disk stars. However, this result will need to be confirmed within another survey with a more accurate metallicity scale. Spatially, the candidate metal-poor thin disk as well as the intermediate metallicity thin disk is located in the outer Galaxy compared to the more metal-rich thin disk component located in the inner Galaxy consistent with a negative metallicity gradient which is observed in the literature \citep[e.g.][]{Cheng2012}.

\item Finally, in section \ref{sec:chemtag} we put forward a powerful chemically labeling approach to separate the chemically distinct canonical thin disk, thick disk+halo, and accreted halo components using the \afenospace, [Al/Fe], [C+N/Fe], and [Mg/Mn] abundance ratios. For consistency in Figure \ref{fig:alphamet_chemtag} we plot our separated components on the standard \afenospace-metallicity plane. Finally, we verified this method by confirming that the kinematics of the resulting populations matched the expected distribution in $l$-GRV space (Figure \ref{fig:lGRV_chemtag}).

\end{itemize}

Ultimately the precise proper motions and distances from Gaia will allow us to study the full spatial position and dynamics of the stars from each of the components to compliment the chemical evolution analysis done here and help solidify the existence of the accreted halo and candidate metal-poor thin disk. Future surveys such as 4MOST \citep{de_jong2012} and ongoing surveys such as the Gaia-ESO survey \citep{Gilmore2012} and GALAH \citep{De_silva2015} will be able to further study the chemical peculiarities of the accreted halo in a statistical way. With full space motions, positions and chemistry of a large sample of stars in hand, we may not only be able to study the accreted \apoor sequence but also confirm the existence of the very metal-poor thin disk and chemical continuity between the halo and thick disk.

\section*{Acknowledgements}
We greatly thank the referee whose comment vastly improved the communication of our results in this paper. K.H. is funded by the British Marshall Scholarship program and the King's College, Cambridge Studentship. This work was partly supported by the European Union FP7 programme through ERC grant number 320360. The SDSS-III/APOGEE Survey made this study possible.

\bibliography{Apogeeletterbib}
\label{lastpage}

\end{document}